\documentclass{emulateapj}	%EmulateApJchange

\tabletypesize{\scriptsize}

\newcommand{\KICRotationPeriod}{22.9}
\newcommand{\KeplerSeventeenRotationPeriod}{12.3}
\newcommand{\HATPElevenRotationPeriod}{30.0}
\newcommand{\KeplerSevenEightRotationPeriod}{12.8}
\newcommand{\KICRotationPeriodShort}{21.9}
\newcommand{\KICRotationPeriodLong}{23.9}
\newcommand{\BootstrapScrambleIterations}{1000}
\newcommand{\TransitTimingLimitSeconds}{70}	
\newcommand{\TransitTimingLimitTwiceSeconds}{140}
\newcommand{\TransitTimingLimitFirstHalfSeconds}{160}
\newcommand{\TransitTimingVariationsKawaharaSeconds}{205}	
\newcommand{\TransitTimingVariationsOccultedSpotsSeconds}{240}	
\shorttitle{The Relation between KIC 1255b's Transit Depths \& the Stellar Rotational Period} 
\shortauthors{Croll et al.}

\begin{document}

\title{The Relation between the Transit Depths of KIC 12557548b \& the Stellar Rotation Period}
\author{Bryce Croll\altaffilmark{1},\altaffilmark{2}
Saul Rappaport\altaffilmark{2},
Alan M. Levine\altaffilmark{2}
}

\altaffiltext{1}{5525 Olund Road, Abbotsford, B.C. Canada}

\altaffiltext{2}{Kavli Institute for Astrophysics and Space Research, Massachusetts Institute
of Technology, Cambridge, MA 02139, USA; croll@space.mit.edu}

\begin{abstract}

Kawahara and collaborators analyzed the transits of the candidate disintegrating Mercury-mass planet
KIC 12557548b and suggested that the transit depths were correlated with the phase of the stellar rotation.
We analyze the transit depths of KIC 12557548b and confirm that there is indeed a robust, statistically
significant signal in the transit depths at the rotation period of the spotted host star.
This signal is more prominent in the first-half of the {\it Kepler} data, and 
is not due to leakage of the rotating spot signal into our measurement of the transit depths, or due
to unocculted starspots.
We investigate the suggestion that this signal could be due to an active region on the star, emitting enhanced
ultraviolet or X-ray radiation leading to an increased mass loss rate of the planet;
we confirm that such a scenario could cause both modulation of the transit depths of KIC 12557548b, and 
small enough transit-timing variations that they might not be detected in the {\it Kepler} data.
Our preferred explanation for the fact that the transit depths of KIC 12557548b are modulated
with the stellar rotation phase is that
the candidate transiting planet is occulting starspots on this highly spotted star; such a scenario
could cause transit depth variations as large as have been observed, and cause 
transit-timing variations small enough that they are arguably consistent with the {\it Kepler} data.

% Based on a simple model we find it unlikely that the modulated
% transit depth signal is due to enhanced mass loss rate as the planet moves
% through a region of enhanced radiation on the host star.

\end{abstract}


%Need to do a plot that compares the depth with aperture radii

\section{Introduction}

The unprecedented precision of the {\it Kepler} space telescope's \citep{Borucki09,Koch10} photometry
has resulted in the discovery of a wealth of 
intriguing exoplanet systems; one such system is the candidate disintegrating 
Mercury-mass exoplanet KIC 12557548b \citep{Rappaport12}.
The {\it Kepler} photometry of the K-dwarf 
host star KIC 12557548 (hereafter referred to as KIC 1255, while the exoplanet will be denoted KIC 1255b)
displays dips that repeat every $\sim$15.7 hours
that vary in depth
from being undetectably shallow to as large as 1.3\% of the host star's flux.
Intriguingly, KIC 1255b's transit profile displays a sharp ingress, followed
by a gradual egress; for these reasons, \citet{Rappaport12} suggested that KIC 1255b may feature a long
cometary tail streaming behind the candidate planet.
In this scenario, KIC 1255b's transits are believed to be
caused by light being scattered out of the line-of-sight from small particles\footnote{Originally thought to be 
sub-micron-sized particles, although \citet{CrollKIC} suggest the largest particles
in the cometary tail should be $\sim$0.5 $\mu m$ or larger.}
trailing behind the planet.
The variable transit depths were suggested to
be due to different amounts of material being ejected from the planet for each orbit.
% No discernible pattern was presented to explain the variations in depth from transit to transit.
 
Such an interesting system has quickly resulted in a number of follow-up observations and reanalyses
of the original {\it Kepler} data. 
These include reanalyses of the {\it Kepler}-data largely supporting
the disintegrating planet scenario \citep{Brogi12,Budaj13,vanWerkhoven14},
theoretical efforts exploring the disintegration of the planet \citep{PerezBeckerChiang13},
and comparisons of the depth of the transits of KIC 1255b obtained in the optical with {\it Kepler},
to those obtained in the near-infrared with the Canada-France-Hawaii Telescope \citep{CrollKIC}.
Perhaps the biggest unresolved issue presented by all the follow-up efforts,
has been the suggestion that KIC 1255b's transit depth variations are correlated
with the phase of the stellar rotation period \citep{Kawahara13}.

In the original \citet{Rappaport12} paper the authors were not able to identify a pattern to explain
the dramatic variability
of KIC 1255b's transit depths.
\citet{Kawahara13} analyzed KIC 1255b's transit depths and presented evidence that 
the observed transit depths
were modulated at the stellar rotation period ($P_{rot}$ $\sim$ \KICRotationPeriod \ d);
the authors found that the transits were on average 30\%
deeper during one phase of the stellar rotation than another.
\citet{Kawahara13} went on to suggest that the 
presumed variable mass loss rate of the planet may therefore be a byproduct of the stellar activity,
and related to an active longitude or starspot group on the star. 
In the \citet{Kawahara13} scenario, when the planet passes
over this active region it may be subjected to increased ultraviolet and X-ray radiation,
or some sort of star-planet interaction arising from magnetic reconnection, leading to an increased mass-loss rate.
However, one problem that quickly arises with the proposed explanation is that
even though an active longitude on the star would only be visible to an observer on the Earth once per stellar
rotation period, the planet would pass over this active longitude each and every orbit.
The travel time for dust grains to pass from the
planet to the end of the cometary tail is expected to be of the order
of an orbital period, while an estimate for the sublimation lifetime
of the grains is several hours \citep{Rappaport12};
therefore even if there were an active longitude on the star
causing increased disintegration of the planet -- either from blasting
the planet with intense ultraviolet or X-ray radiation, or subjecting it to some
sort of magnetic star-planet interaction -- 
the effects would be spread over a significant duration of the 
extremely short $\sim$15.7 hour orbital period of the planet.

 More benign explanations for the observed relation include the effects of occulted and unocculted
starspots, or leakage from the significant rotational starspot modulation of KIC 1255 causing
mis-estimates of KIC 1255b's transit depths.
KIC 1255 displays obvious rotational 
modulation with peak to peak variations between 
$\sim$1\% and $\sim$5\% of the observed stellar flux with a rotation period of $\sim$\KICRotationPeriod \ d.
Unocculted and occulted spots are known to bias the measurements of transit depths (e.g. \citealt{Czela09}; \citealt{Carter11});
furthermore, if the significant rotational modulation leaks into measurements of the planet's
transit depths, the reported signal could simply be an artefact.
\citet{Kawahara13} considered the possibility that unocculted spots could introduce a correlation between
KIC 1255b's transit depths and the rotation period, but suggested that the 30\% transit depth signal
was far greater than would be expected for unocculted spots.
% They did not consider occulted starspots.

 In this paper we attempt to verify whether the claimed KIC 1255 
Transit Depth - Rotation Modulation 
Signal (hereafter referred to as the TDRM signal)
is genuine, and if it is due to something more astrophysically interesting
than simply occulted and unocculted spots.
In Section \ref{SecAnalysis} we demonstrate that the TDRM signal is not due to an artefact,
and is statistically significant; however, we show that other spotted stars also display a similar signal due to occulted spots.
We also present a transit-timing analysis of the transits of KIC 1255b.
In Section \ref{SecExplain} we examine three scenarios for the TDRM signal. We rule out unocculted spots, 
and demonstrate
that the scenario suggested by \citet{Kawahara13} -- of an active region with enhanced ultraviolet radiation leading
to an increased mass loss rate -- could naturally cause the transit depth variations that are observed.
However, we argue that occulted spots are arguably the simplest explanation presented to date for the TDRM signal.

% \citet{CrollKIC} presented multiwavelength observations of KIC 1255b from the optical with {\it Kepler} to the near-infrared
% with the Canada-France-Hawaii Telescope that suggested that the largest particles in the cometary
% tail trailing KIC 1255b must be at least $\sim$0.5 microns in size or larger.

% Our original belief was that the claimed \citet{Kawahara13} relation was simply an artefact of inadequate removal
% from the transit depths of the obvious rotational
% modulation displayed by KIC 1255. 

\section{Analysis of the Transit Depth Rotational Modulation Signal}
\label{SecAnalysis}

\subsection{Basic analysis}
\label{SecKawahara}

%EMULATEAPJCHANGE 0.40 to 0.42
\begin{figure*}
\centering
\includegraphics[scale=0.47, angle = 270]{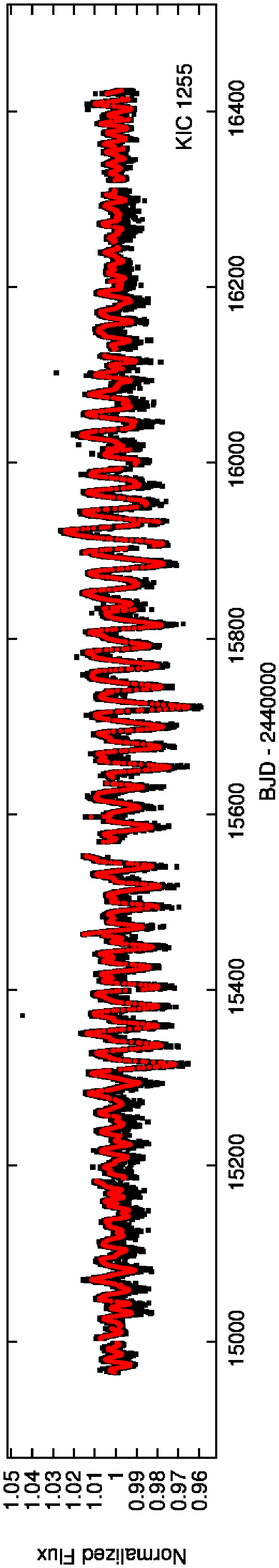}
\includegraphics[scale=0.47, angle = 270]{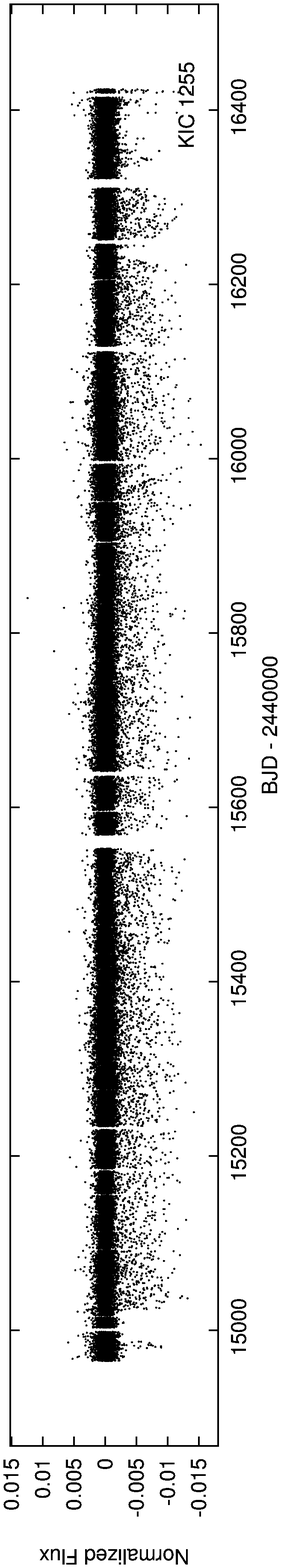}
\includegraphics[scale=0.42, angle = 270]{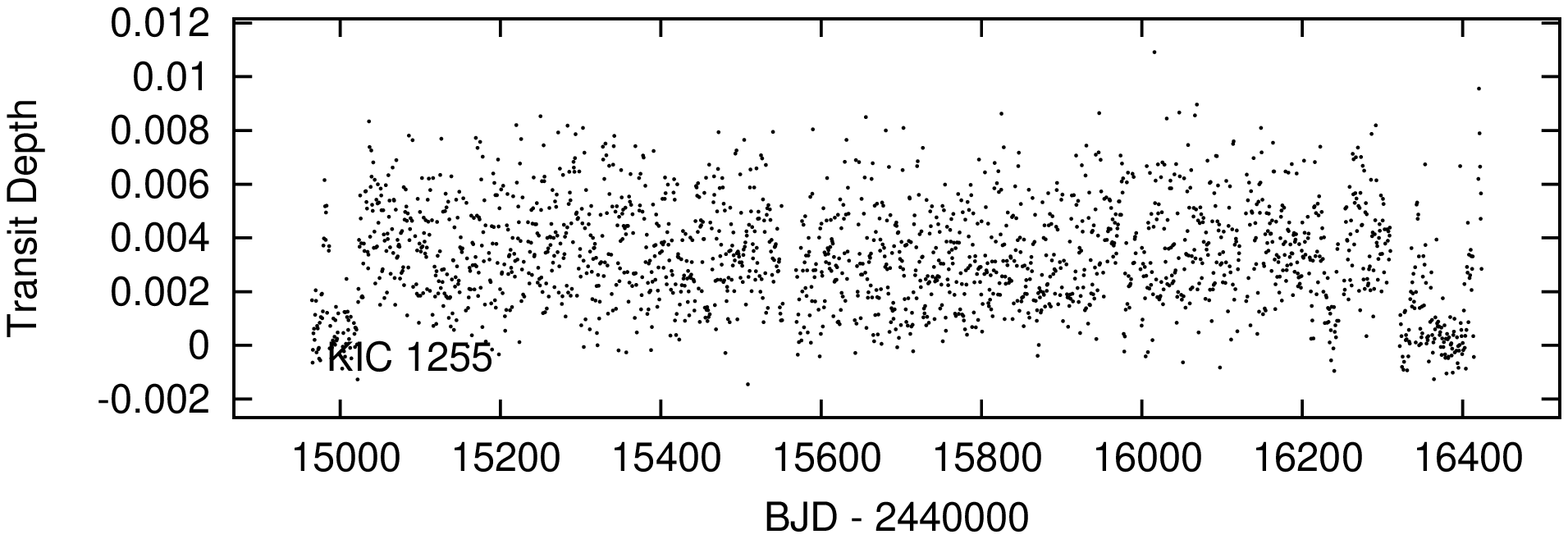}
\includegraphics[scale=0.42, angle = 270]{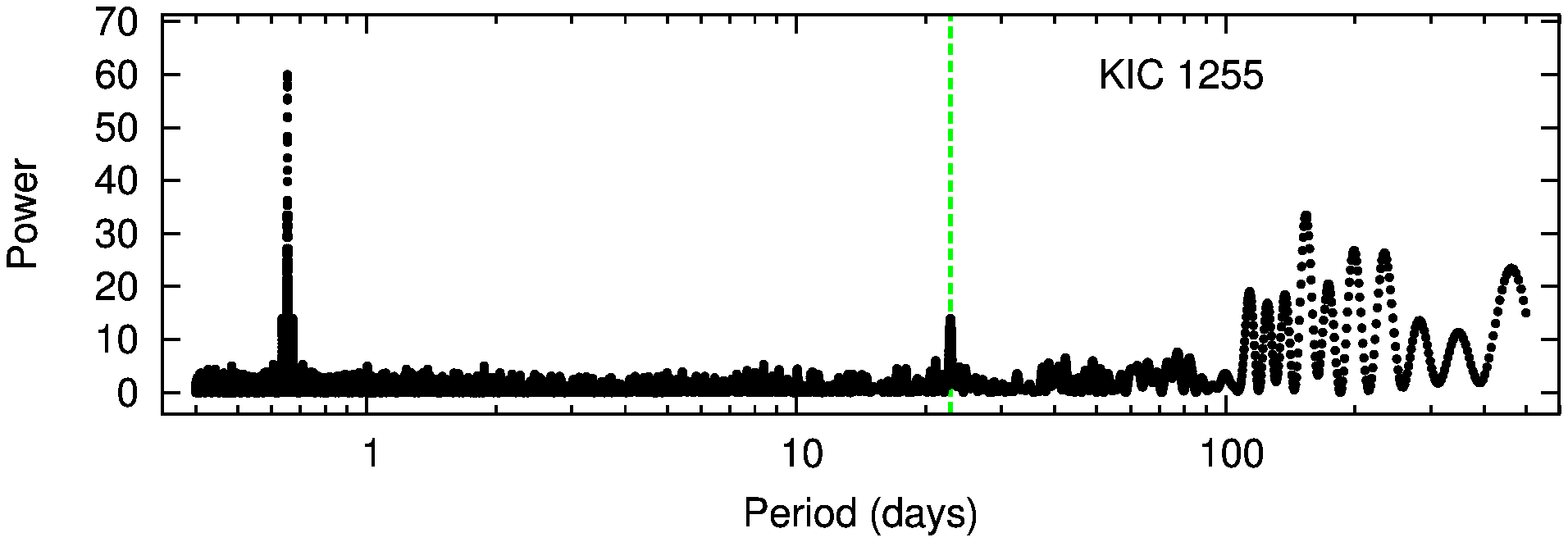}
\caption[KICKepler]
	{
		Top panel: The {\it Kepler} long cadence photometry of KIC 1255 (black points);
		the red points display the values used in the cubic spline fit
		to subtract out KIC 1255's obvious rotational modulation.
		Middle panel: The {\it Kepler} long cadence photometry after the subtraction of the spline fit
		to remove the obvious rotational modulation. The transits 
		every $\sim$15.7 hours of KIC 1255b are obvious as the sea
		of points just below the mean of the light curve.
		Bottom left panel: The transit depths of the {\it Kepler} long cadence photometry of KIC 1255b.
		Bottom right panel: The Lomb Scargle periodogram of the transits of KIC 1255b. 
		The vertical dotted green line displays the $\sim$\KICRotationPeriod \ d rotation period of KIC 1255.
 	}
\label{FigKIC}
\end{figure*}

We first largely repeat the analysis presented in \citet{Kawahara13} using our own techniques.
We start by analyzing the {\it Kepler} long cadence ($\sim$29.4 minute sampling) photometry of KIC 1255 
(quarters 1-17)\footnote{Our 
analysis features two quarters of data that were not available at the time of the \citet{Kawahara13} analysis.}.
We utilize the pre-search data conditioning
simple aperture photometry (PDCSAP; \citealt{Smith12}; \citealt{Stumpe12}) of this star.
We normalize the light curve for each {\it Kepler} quarter to the median flux observed in that quarter.
The success of our analysis depends on ensuring that the transit depths of KIC 1255b are accurately measured despite 
KIC 1255's obvious rotational flux modulation.
In order to ensure that this is the case, we remove the rotational modulation by subtracting out a cubic 
spline fit.
To ensure that the asymmetrical transit of KIC 1255b does not affect our removal of the rotational modulation,
we cut out all data in the transit before calculating our cubic spline.
That is, we phase the data to the orbital period of the
candidate planet (where phase, $\phi$=0.5 denotes the midpoint of
the transit), and
cut out all data between
phases of $\phi_{mincut}$ = 0.4 to $\phi_{maxcut}$=0.7. 
We then bin the data every $\sim$10 hr, and use these data to calculate our cubic spline,
and remove the obvious rotational modulation.
We apply a 10$\sigma$ cut to the spline-corrected data to remove obvious outliers.
Relevant analysis parameters are summarized in Table \ref{TableAnalysis}.
The photometric data of KIC 1255 before and after the removal of
the rotational modulation are displayed in the top panels of Figure \ref{FigKIC}.

\begin{deluxetable}{cccccc}
\tablecaption{Analysis Parameters}
\tabletypesize{\scriptsize}
\tablehead{
\colhead{Star}  & \colhead{$\phi_{mincut}$}	& \colhead{$\phi_{maxcut}$}	& \colhead{sigma-cut}	& \colhead{$P_{orbit}$} & \colhead{$P_{rot}$}	\\	
\colhead{}  	& \colhead{}			& \colhead{}			& \colhead{}		& \colhead{(d)} 	& \colhead{(d)}	\\	
}
\startdata
KIC 1255	& 0.40				& 0.60 				& 10$\sigma$		& $\sim$0.65 		& $\sim$\KICRotationPeriod 		\\
Kepler-17	& 0.45				& 0.55 				& n/a			& $\sim$1.49 		& $\sim$\KeplerSeventeenRotationPeriod 	\\
HAT-P-11	& 0.48				& 0.52 				& n/a			& $\sim$4.89 		& $\sim$\HATPElevenRotationPeriod 	\\
Kepler-78	& 0.40				& 0.60 				& n/a			& $\sim$0.36 		& $\sim$\KeplerSevenEightRotationPeriod	\\
\enddata
% \tablenotetext{a}{We fix $t_{transit}$ to the predicted mid-point of the transit for this analysis, due to the fact we are unable to detect the transit on this occasion.}
\label{TableAnalysis}
\end{deluxetable}

\citet{Kawahara13} measured the transit depths of KIC 1255b by averaging the three long cadence {\it Kepler}
points nearest to the mid-point of the transit, and we do likewise.
This method has the obvious drawback of imprinting the 
frequency of the long cadence photometry into the transit signal, but as we are interested in a signal
with a much longer period than the $\sim$29.4 minute cadence of long cadence photometry, this effect
does not seriously impact our science goals.
We have experimented with formally fitting the transit depths of KIC 1255b, as we do in \citet{CrollKIC}, and
we've confirmed that the results are similar to the method we use here.

%EMULATEAPJCHANGE, 0.42 to 0.40
\begin{figure*}
\centering
\includegraphics[scale=0.42, angle = 270]{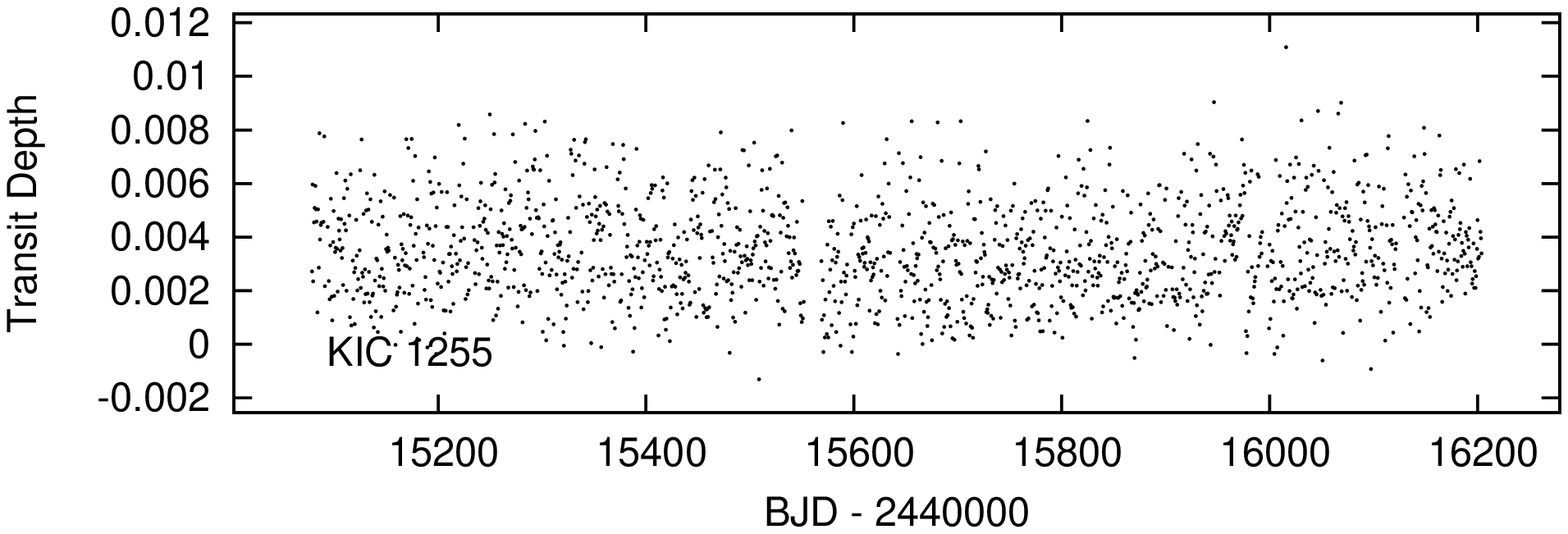}
\includegraphics[scale=0.42, angle = 270]{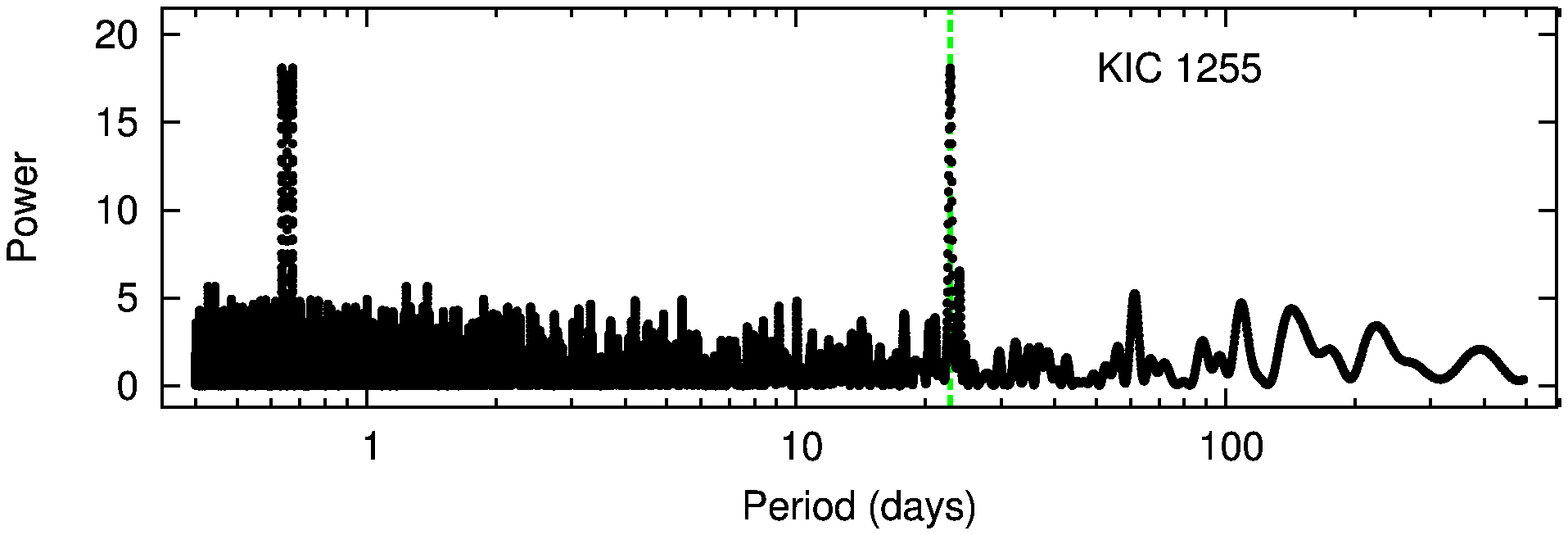}
\caption[KICKepler]
	{
		Left panel: The transit depths of the {\it Kepler} long cadence photometry of KIC 1255b,
		excluding the quiescent periods, by cutting all data
		before BJD = 2455078.0 and after BJD = 2456204.3.
		Right panel: The associated Lomb Scargle periodogram of the transits of KIC 1255b. 
		The vertical dotted green line displays the $\sim$\KICRotationPeriod \ d rotation period of KIC 1255.
 	}
\label{FigKICExcludeShallow}
\end{figure*}

We then take a Lomb Scargle periodogram \citep{Lomb76,Scargle82} of the KIC 1255b transit depths (bottom right panel of Figure \ref{FigKIC}).
Similarly to \citet{Kawahara13} we notice a modest peak in the periodogram near the rotation
period of the star\footnote{We determine the rotation period of the star by identifying the maximum peak in the Lomb Scargle periodogram of
the {\it Kepler} long cadence data, once the transit data has been removed from $\phi_{mincut}$ to $\phi_{maxcut}$,
as given in Table \ref{TableAnalysis}.}. 
The peak at $\sim$0.65 d is simply due to the orbital period of the planet.
% short periods in our Long periodogram of the transit depths results from the cadence of the long cadence photometry, which has been imprinted into our transit depths due to our method of determining the transit depths mentioned above.
There are more prominent peaks in the periodogram at longer periods than the rotation period;
these periods were also mentioned in \citet{Kawahara13}. These longer period peaks are largely due to the 
quiescent periods at the start and the end of the data-set when the KIC 1255b transit depths are virtually
undetectable (these quiescent periods were noted in \citealt{vanWerkhoven14}).
If we repeat the analysis excluding these points -- by excluding
data before BJD = 2455078.0 and data after BJD = 2456204.3 -- the results are shown in Figure \ref{FigKICExcludeShallow}.
There does appear to be a prominent peak in the periodogram near the rotation period of the star.

\subsubsection{Further investigation of the suggested TDRM signal}

\begin{figure}
\centering
\includegraphics[scale=0.42, angle = 270]{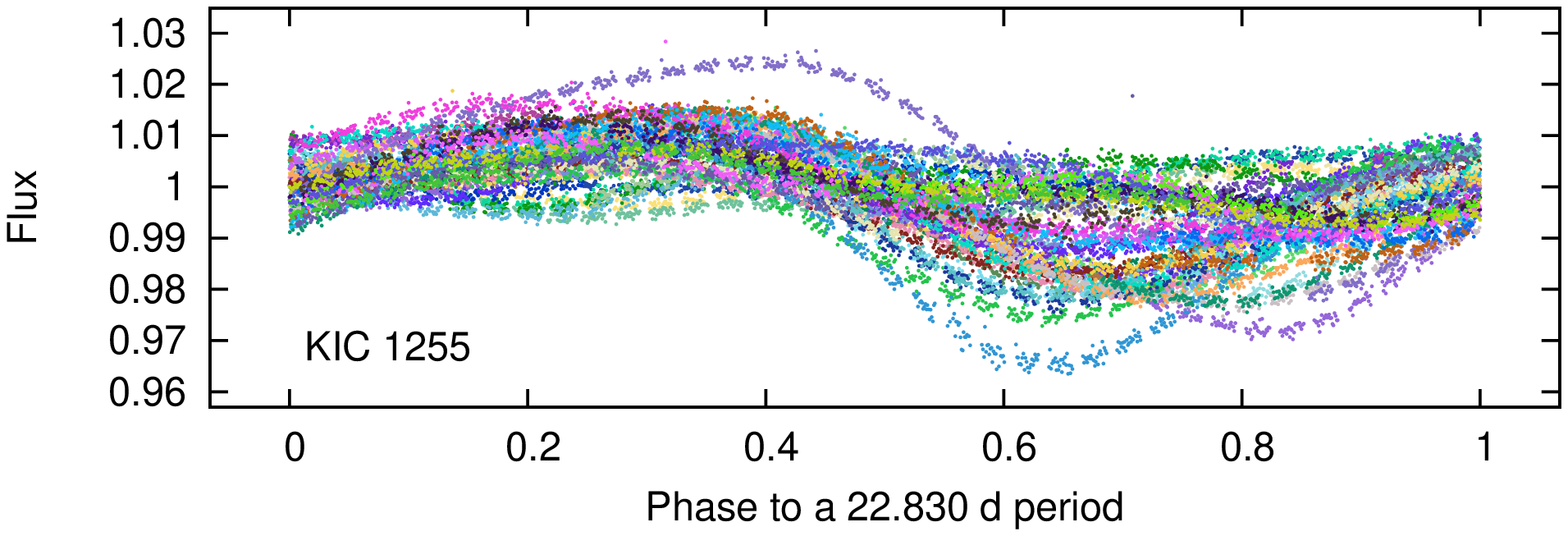}
\includegraphics[scale=0.42, angle = 270]{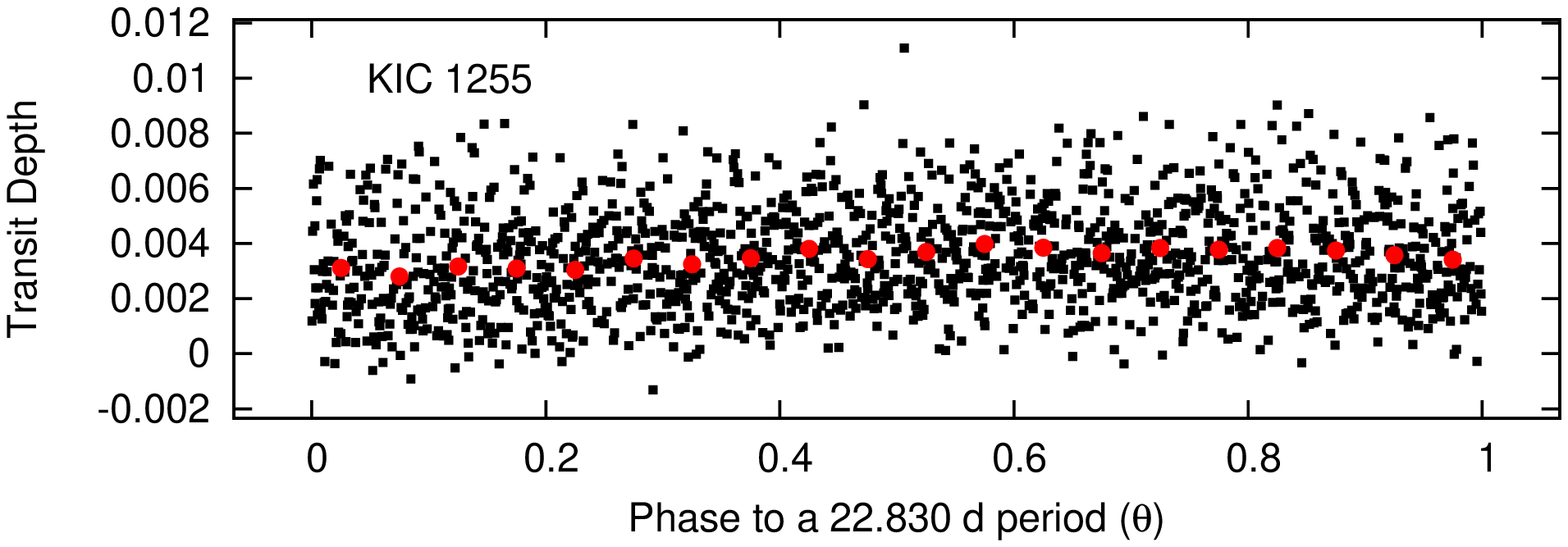}
\includegraphics[scale=0.42, angle = 270]{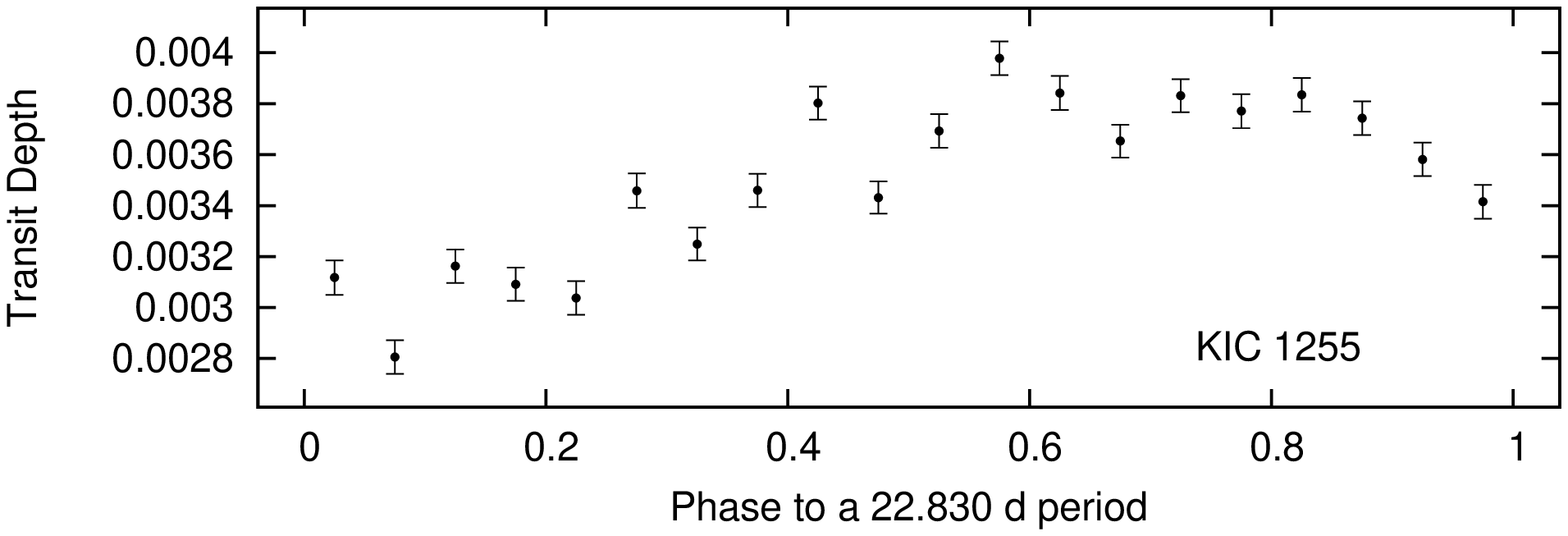}
\caption[KICKepler]
	{
		Top panel: The KIC 1255 {\it Kepler} long cadence photometry with the transits removed
		phased to the rotation period of the star.
		Each rotation period of the star is drawn with a different colour.
		Middle panel: The transit depths phased to the rotation period of the star (black points),
		and the transit depths binned every $\phi$=0.05 in phase (red points).
		Bottom panel: The same binned transit depths phased to the stellar rotation period.
		Although the deepest transits appear to be roughly coincident with the mean minimum of the rotational
		modulation, in fact Figure \ref{FigSpotsLightcurve}
		indicates that there is no well defined rotational modulation flux minimum for this star.
 	}
\label{FigKICPhase}
\end{figure}

\begin{figure*}
\centering
\includegraphics[scale=0.50, angle = 270]{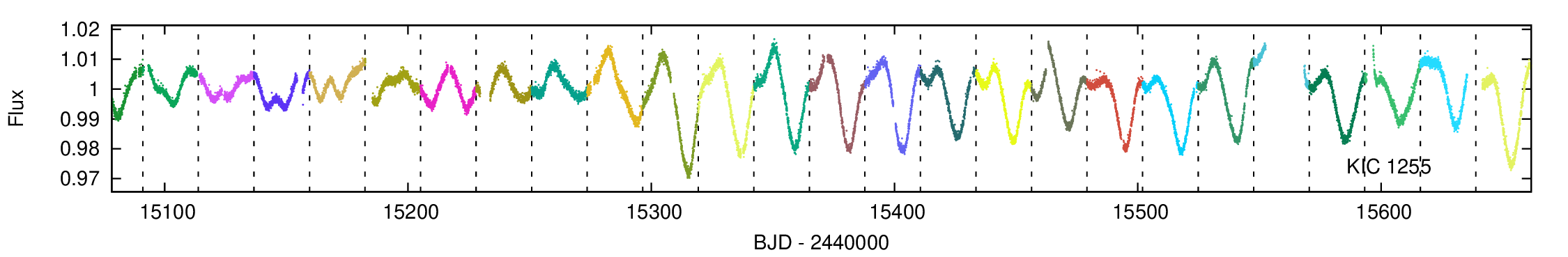}
\includegraphics[scale=0.50, angle = 270]{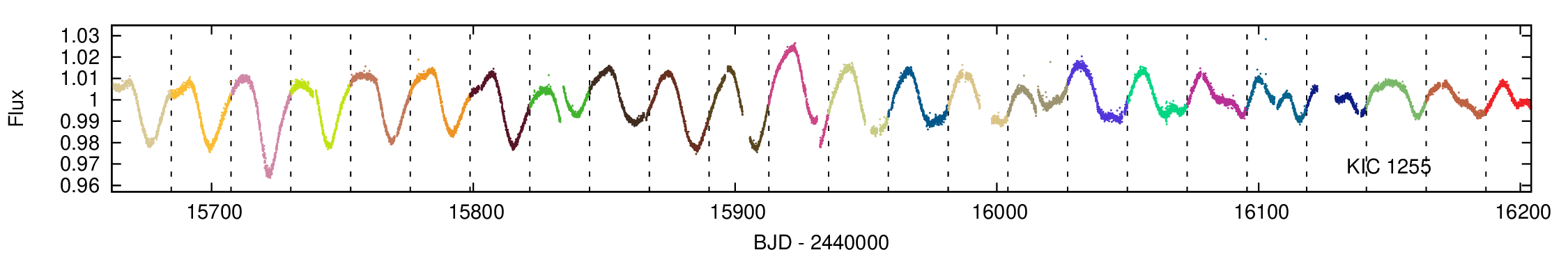}
\caption[]
	{
		The long cadence {\it Kepler} photometry of KIC 1255 with the transits removed.
		The data are split into two plots, with the later dates at bottom.
		Each additional colour, and the vertical dashed lines,
		represent an additional rotation period of the star. The effects of spots
		on the light curve are visible at all phases of the apparent
		stellar rotation period, and therefore at all longitudes of the star if we are viewing the star edge-on.
 	}
\label{FigSpotsLightcurve}
\end{figure*}

We also phase the transit depths, excluding those data before and after the quiescent periods (Figure \ref{FigKICExcludeShallow}),
to the rotation period of the star as seen in Figure \ref{FigKICPhase}
(we will denote the phase to the stellar rotation period as $\theta$, with $\theta$=0 corresponding to
BJD=2455410.654).	% 2455410.6537769
The transit depths phased to the stellar rotation period reach a maximum of approximately $\sim$0.38 $\pm$ 0.01\%
from $\theta$ = 0.54 - 0.74,
compared to a minimum of approximately $\sim$0.30 $\pm$ 0.01\% at $\theta$ = 0.04 - 0.24.
Therefore, the transit depths are 25\% deeper at one phase of the stellar rotation than another; this compares
with the 30\% signal reported by \citet{Kawahara13} - most likely due to the fact that \citet{Kawahara13} removes the rotational
modulation by dividing through by the flux, rather than subtracting\footnote{In the original 
\citet{Kawahara13} analysis the rotational modulation of the star 
was divided out to produce the flattened light curve using the routine {\it kepflatten}.
Subtracting out, rather than dividing out, the rotational modulation of KIC 1255 is
a significant improvement in our present work compared to the original \citet{Kawahara13} study.
Dividing through by the flux of the star that is displaying rotational modulation, imparts the rotational
modulation signal into the transit depths of that star (as discussed in Section \ref{SecUnocculted}).
Anywhere from $<$1\% to 5\% of the modulation
of the transit depths suggested by \citet{Kawahara13} is likely due to this effect, compared to their
reported 30\% TDRM signal.}.
Although \citet{Kawahara13} suggested that the deepest transit depths were roughly coincident with the minimum flux
of the observed rotational modulation (and thus when the starspots were most visible), we illustrate in the top panel
of Figure \ref{FigKICPhase} and in Figure \ref{FigSpotsLightcurve}
that the starspots on KIC 1255 are highly variable. There is not a well defined, constant flux minimum of the rotational modulation.
The effects of spots appear to be apparent
at all phases of the stellar rotation period; if we are viewing the star in the equatorial plane
then this corresponds to spots being 
visible at all longitudes of the star.

\begin{figure}
\centering
\includegraphics[scale=0.42, angle = 270]{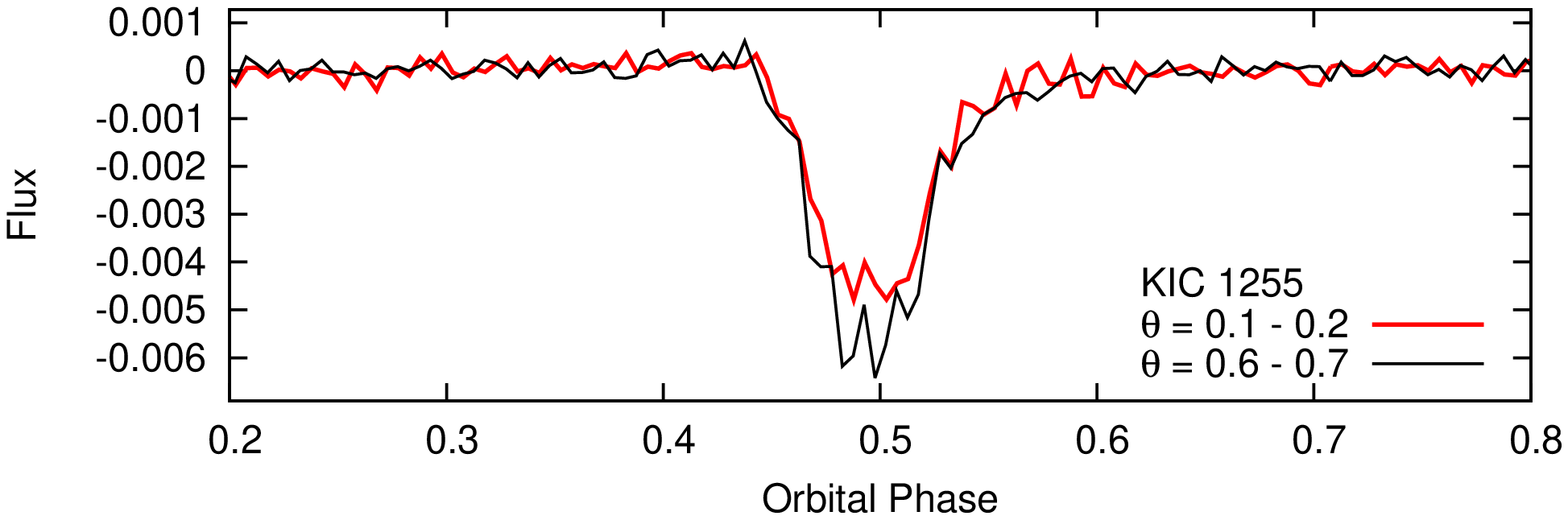}
\caption[]
	{
		The long cadence {\it Kepler} photometry of KIC 1255 phased 
		to the orbital period of the planet.
		In order to inspect transit profile differences
		with the stellar rotation period, we break the data into discrete
		segments with respect to the phase of the 
		stellar rotation period, $\theta$. 
		The mean transit profile for all the {\it Kepler} data from $\theta$ = 0.1 - 0.2,
		corresponding to the shallowest transit depths of the TDRM signal, 
		is displayed with the red solid line, 
		while the mean transit profile for $\theta$ = 0.6 - 0.7, corresponding to the deepest depths
		of the TDRM signal, is displayed with the black solid line.
		The mean transit profile is deeper, but not clearly different in shape for 
		$\theta$ = 0.6 - 0.7, compared to $\theta$ = 0.1 - 0.2.
 	}
\label{FigTransitProfile}
\end{figure}

 We also investigated changes in the shape of the transit profile of KIC 1255b, phased with the stellar
rotation period. We split the data into discrete segments 
every 0.1 in stellar rotation phase, $\theta$.
No obvious transit profile shape changes were observed.
We display the long cadence {\it Kepler} data for approximately the 
deepest ($\theta$ = 0.6 - 0.7) and shallowest ($\theta$ = 0.1 - 0.2)
depths of the TDRM signal in Figure \ref{FigTransitProfile}.
The transit is deeper for $\theta$ = 0.6 - 0.7
compared to $\theta$ = 0.1 - 0.2, but changes in the transit profile
are not evident.

% As discussed above, as we are interested in accurately determining
% the transit depths of KIC 1255b, and  displays obvious rotational modulation, we must ensure that our measurements
% of the transit depths are minimally affected by the rotational modulation.
% To ensure this is the case

\begin{figure*}
\centering
\includegraphics[scale=0.40, angle = 270]{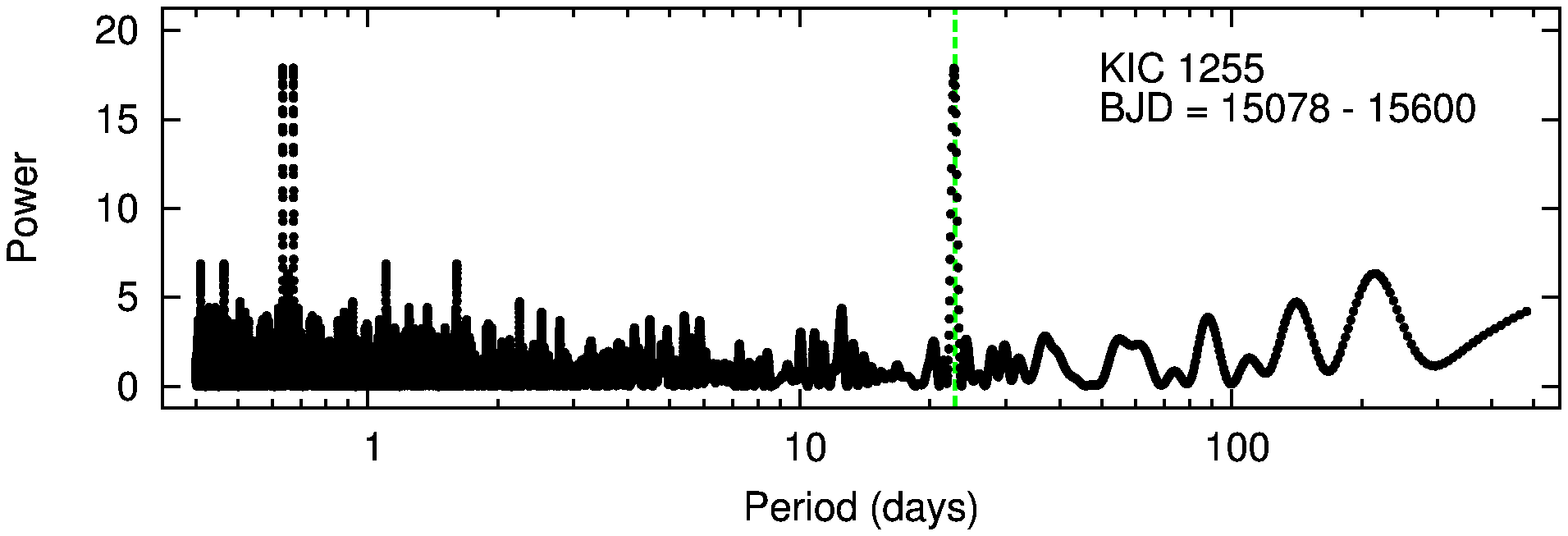}
\includegraphics[scale=0.40, angle = 270]{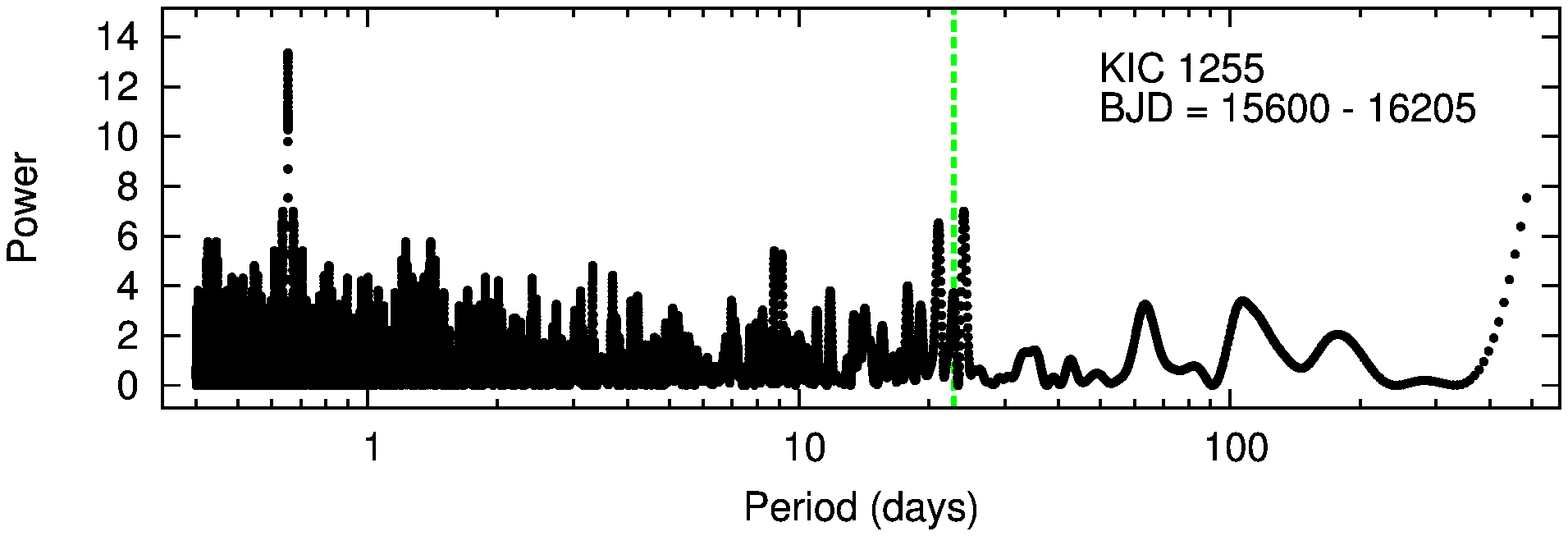}
\includegraphics[scale=0.40, angle = 270]{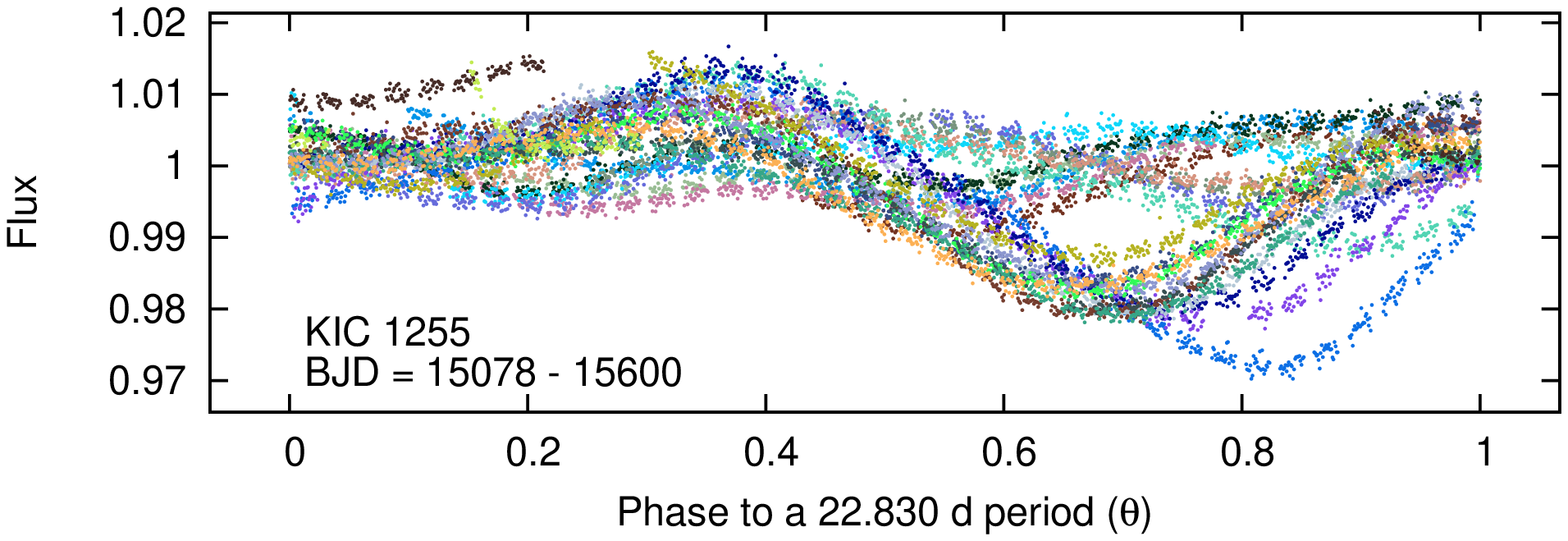}
\includegraphics[scale=0.40, angle = 270]{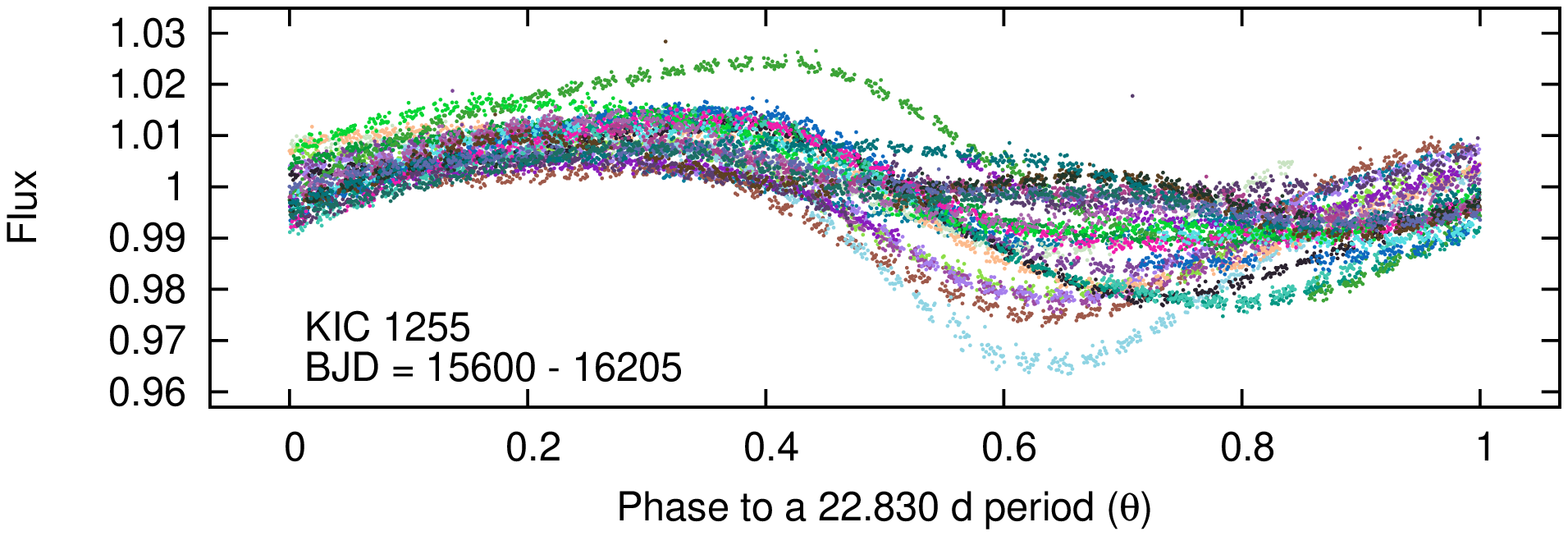}
\includegraphics[scale=0.40, angle = 270]{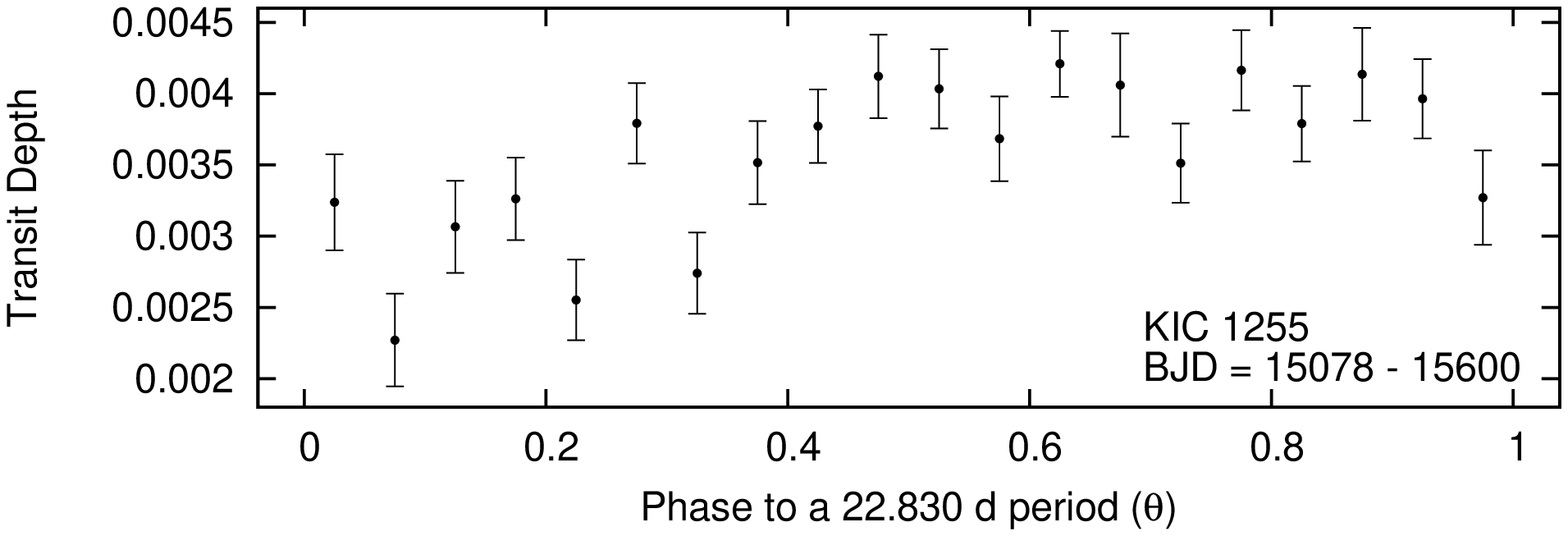}
\includegraphics[scale=0.40, angle = 270]{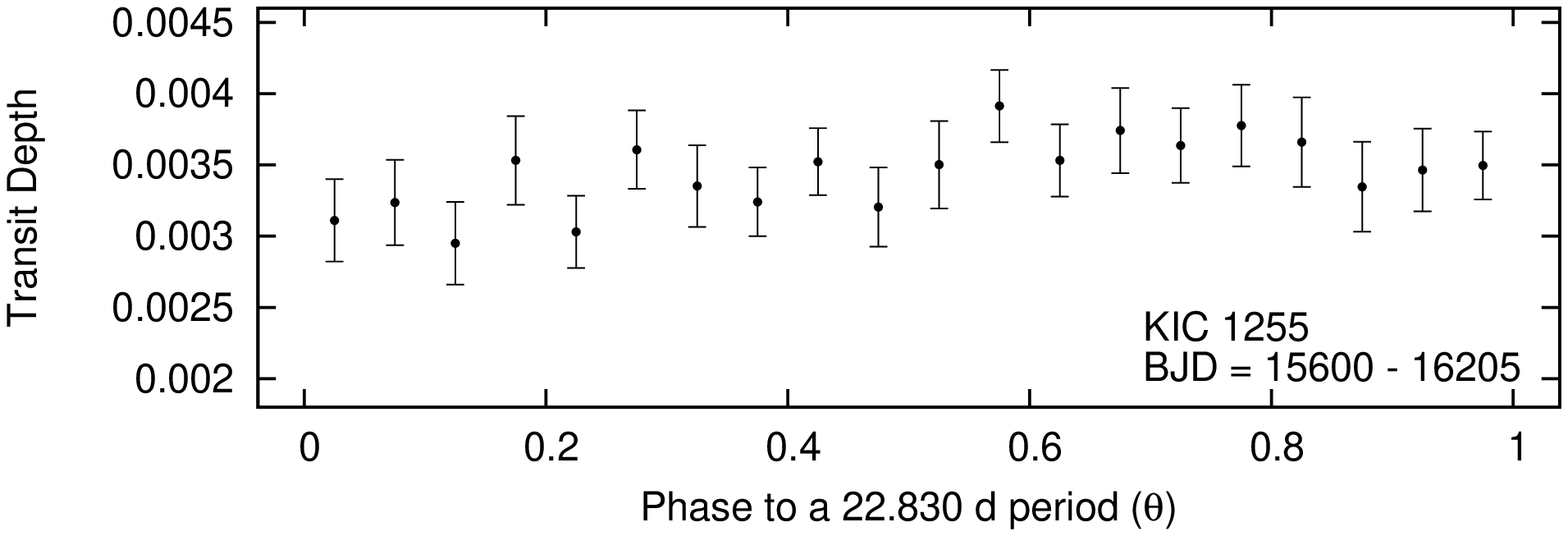}
\caption[]
	{
		Our analysis for the first half of the {\it Kepler} long cadence data excluding the quiescent periods
		(BJD - 2440000 = 15078 - 15600; left panels),
		and for the second half of the {\it Kepler} long cadence data excluding the quiescent periods
		(BJD - 2440000 = 15600 - 16205; right panels).
		The top panels display the Lomb Scargle periodogram of the transit depths, with the stellar
		rotation period denoted by the vertical dotted green line.
		The middle panels show the {\it Kepler} long cadence data
		with the transits removed phased to the rotation period of the star, with
		each additional stellar rotation period plotted in a different colour.
		The bottom panels displays the transit depths phased to the stellar rotation period.
		The TDRM signal is significantly stronger in the early data (left panels), where a spot is visible
		at phase $\theta$$\sim$0.2.
 	}
\label{FigKawaharaSplit}
\end{figure*}

 We have also investigated the strength of the KIC 1255 TDRM signal using different subsets of
the {\it Kepler} data. The TDRM signal is stronger
in the first-half of the data following the quiescent period,
than in the second half. That is, we perform our analysis on the data from
BJD - 2440000 = 15078 - 15600, and from BJD - 2440000 = 15600 - 16205 and display the results
in Figure \ref{FigKawaharaSplit}. The TDRM signal is significantly stronger in the left panels
({\it Kepler} data from BJD - 2440000 = 15078 - 15600), than in the right panels
({\it Kepler} data from BJD - 2440000 = 15600 - 16205). We discuss the possible importance of
this in Section \ref{SecOcculted}.

\subsection{Lessons learned from other transiting planets orbiting spotted stars}
\label{SecStars}

%EMULATEAPJCHANGE, 0.42 to 0.40, for the bottom ones...
\begin{figure*}
\centering
\includegraphics[scale=0.47, angle = 270]{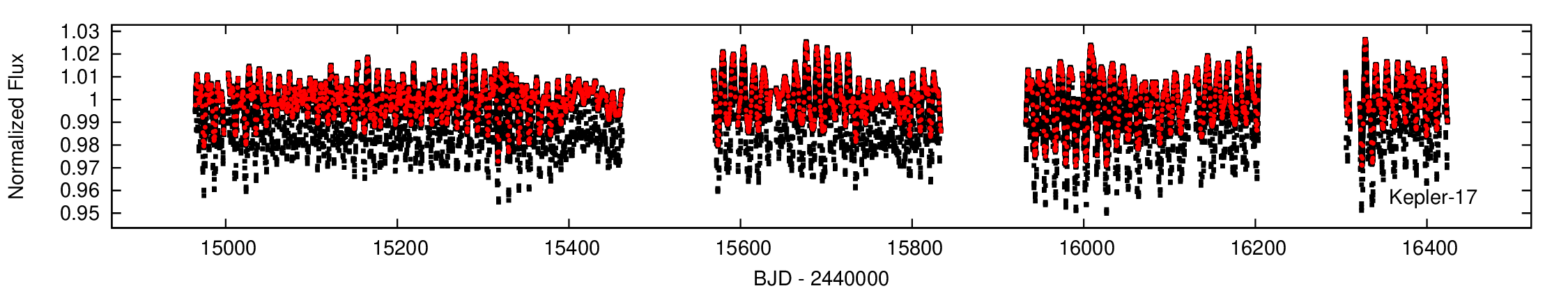}
\includegraphics[scale=0.47, angle = 270]{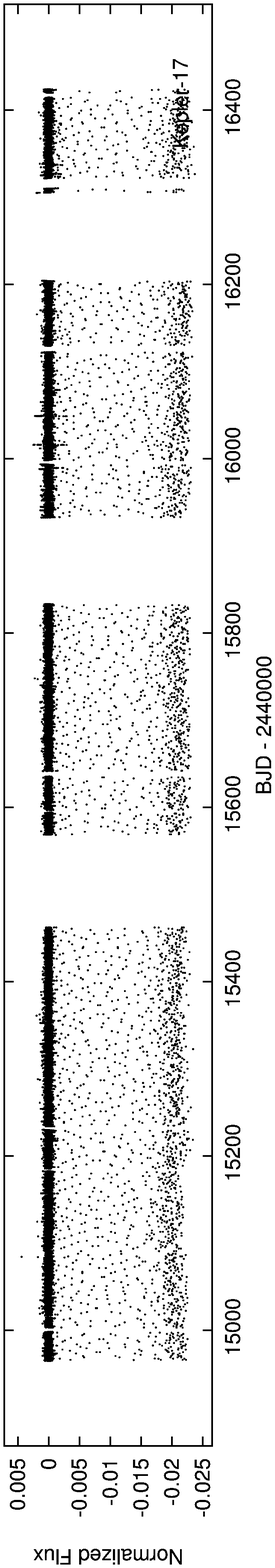}
\includegraphics[scale=0.42, angle = 270]{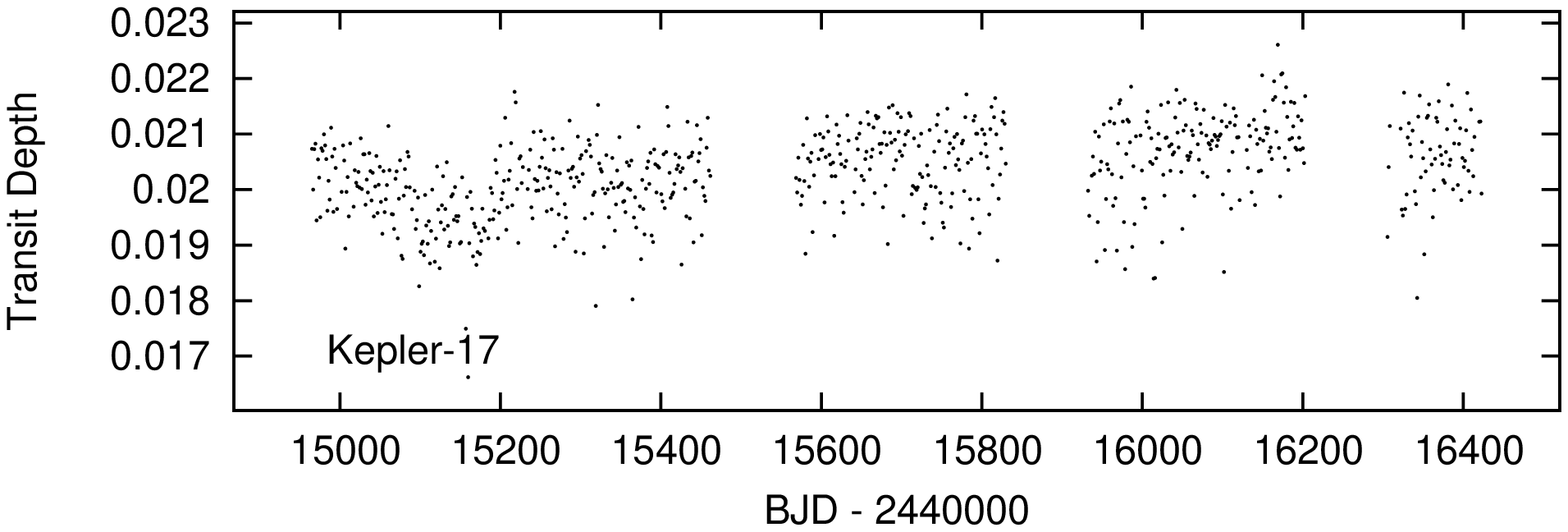}
\includegraphics[scale=0.42, angle = 270]{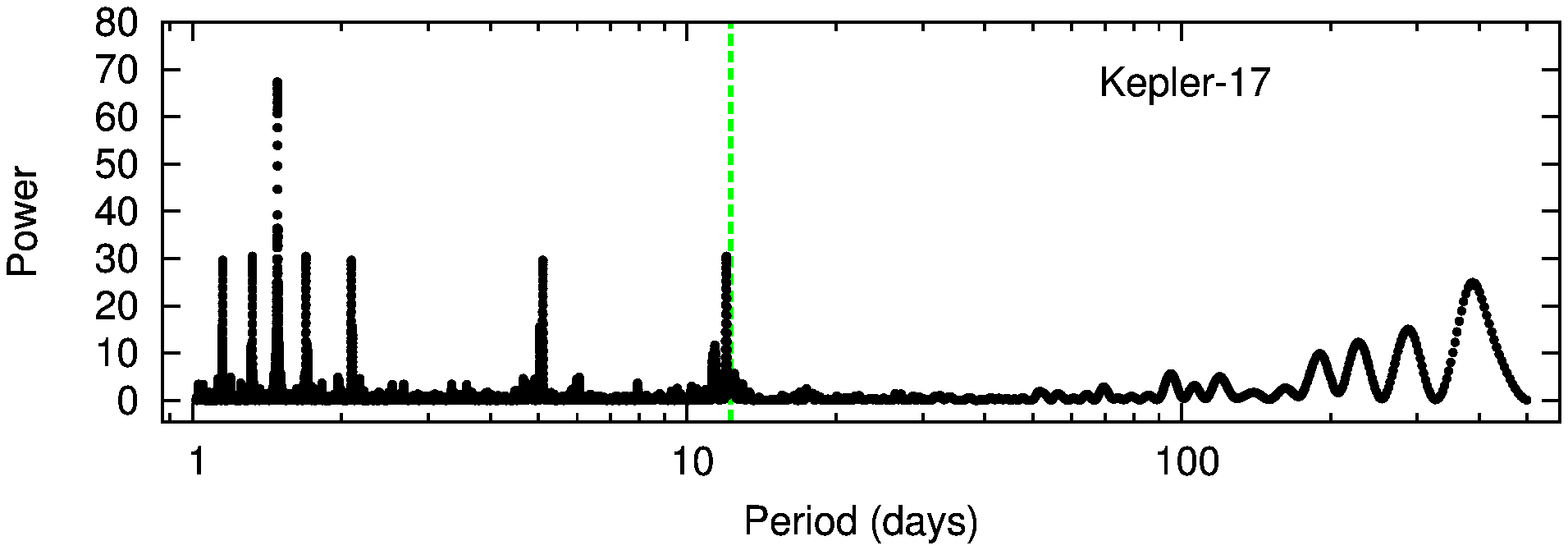}
\caption[]
	{
		The
		Kepler-17
		long cadence photometry (top), the photometry after the removal
		of the rotational modulation (middle), transit-depths (bottom left) and 
		the Lomb Scargle periodogram of the transit depths (bottom right).
		The Figure format is identical to Figure \ref{FigKIC}.
		The vertical dotted green line in the bottom right panel
		indicates the
		$\sim$\KeplerSeventeenRotationPeriod \ d
		rotation period of Kepler-17.
 	}
\label{FigKepler17}
\end{figure*}

%EMULATEAPJCHANGE, 0.42 to 0.40, for the bottom ones...
\begin{figure*}
\centering
\includegraphics[scale=0.47, angle = 270]{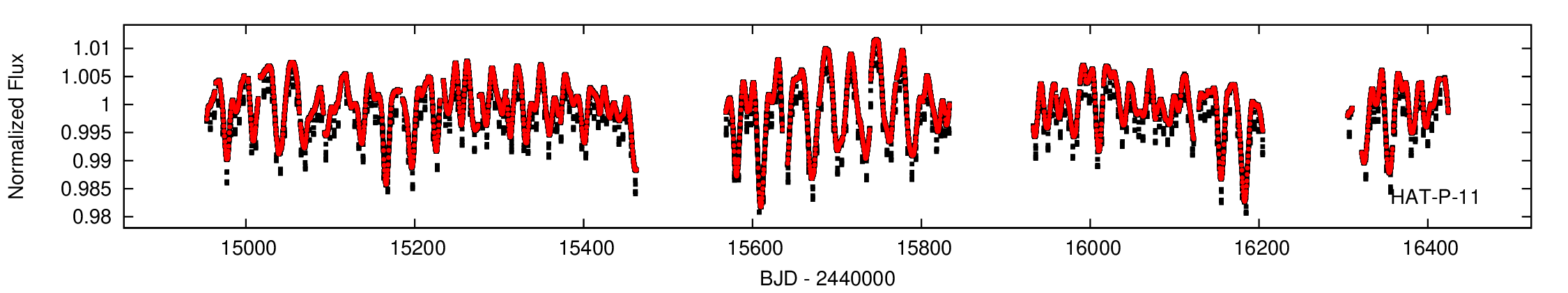}
\includegraphics[scale=0.47, angle = 270]{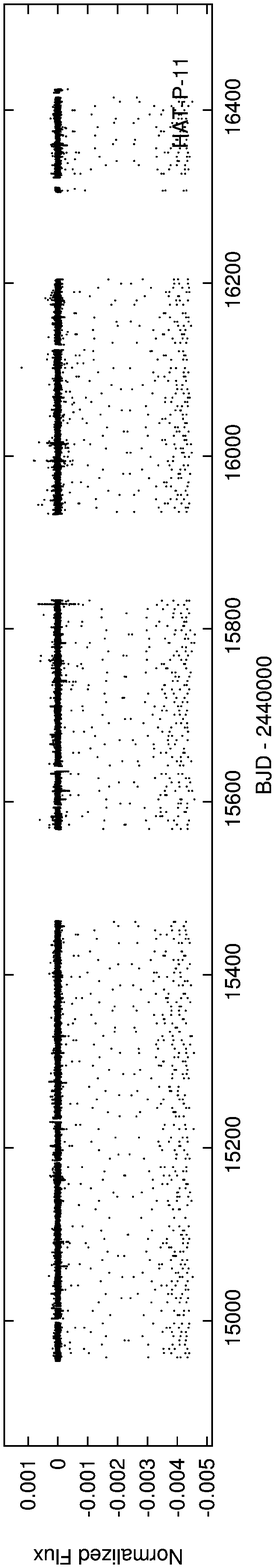}
\includegraphics[scale=0.42, angle = 270]{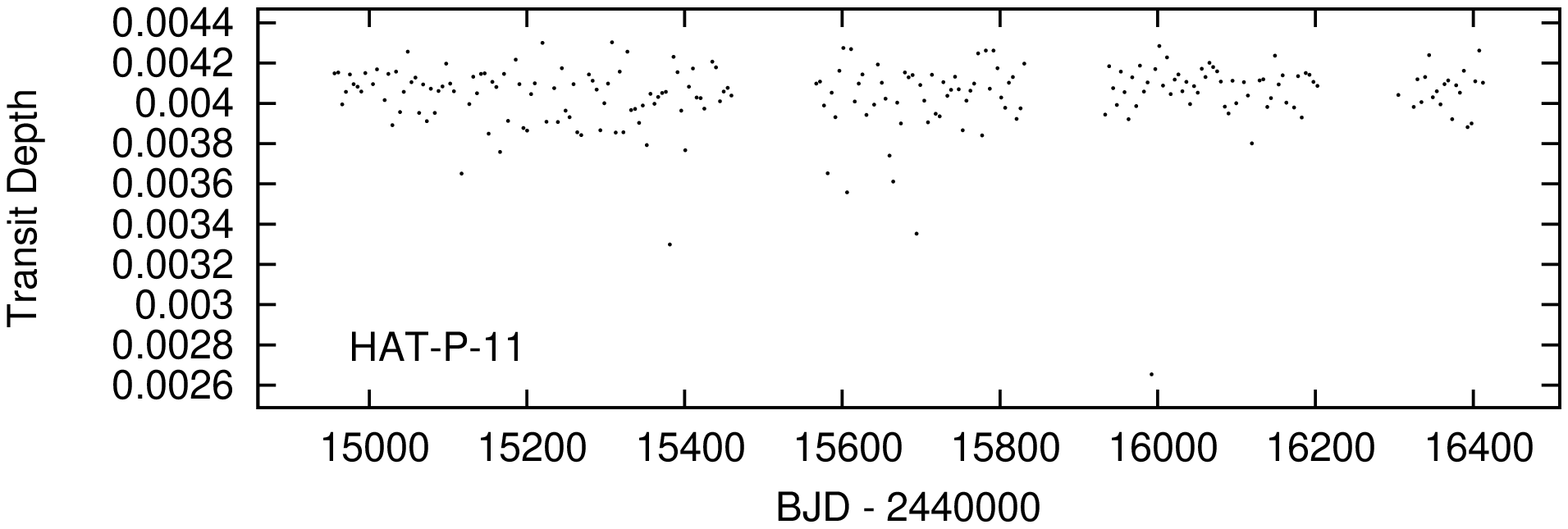}
\includegraphics[scale=0.42, angle = 270]{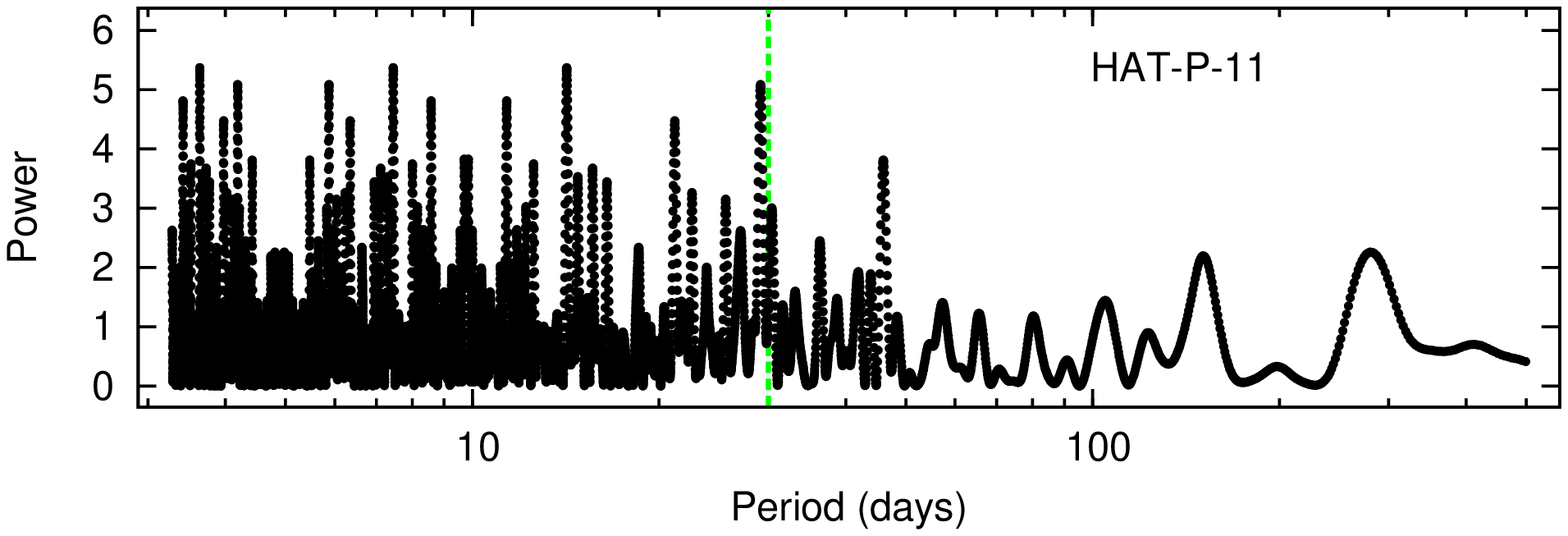}
\caption[]
	{
		The
		HAT-P-11
		long cadence photometry (top), the photometry after the removal
		of the rotational modulation (middle), transit-depths (bottom left) and 
		the Lomb Scargle periodogram of the transit depths (bottom right).
		The Figure format is identical to Figure \ref{FigKIC}.
		The vertical dotted green line in the bottom right panel
		indicates the
		$\sim$\HATPElevenRotationPeriod \ d rotation period of HAT-P-11.
 	}
\label{FigHATP11}
\end{figure*}

We also apply our analysis to the photometry of several other spotted stars hosting transiting planets
that have been observed with {\it Kepler}.
Stars were chosen 
that display significant rotational modulation,
that host a short-period transiting planet,
and that have stellar rotation period that is much longer than the orbital
period of the planet.
These stars include: Kepler-17 \citep{Desert11}, 
Kepler-78 \citep{SanchisOjeda13}, and HAT-P-11 \citep{Bakos10}.
The stars Kepler-17 \citep{Desert11} and HAT-P-11 \citep{Deming11,SanchisOjeda11} 
have been observed to display obvious occulted spots during transit.
As Kepler-17 appears to be spin-orbit aligned, and there is a close commensurability of the orbital and rotation
periods ($\sim$8 orbital periods to each rotation period), 
the spots of Kepler-17 appear to repeat every few transits, corresponding to a rotation period of
the star \citep{Desert11}.
As these occulted spots will result in a slight mis-estimation of the transit depth, one would expect
that the transit depth signal would be modulated by the rotation period of the star due to these occulted spots. 
Similarly, HAT-P-11, although it is spin-orbit misaligned \citep{Deming11,SanchisOjeda11}, has spots that appear to be relatively long-lived, 
and thus occulted spots may reappear a rotation period later, causing a possible TDRM signal.
Unocculted starspots can also bias measurements of the transit depths, but our technique of subtracting out
the rotational modulation should largely remove this bias, as discussed in Section \ref{SecUnocculted}.

\begin{figure*}
\centering
\includegraphics[scale=0.47, angle = 270]{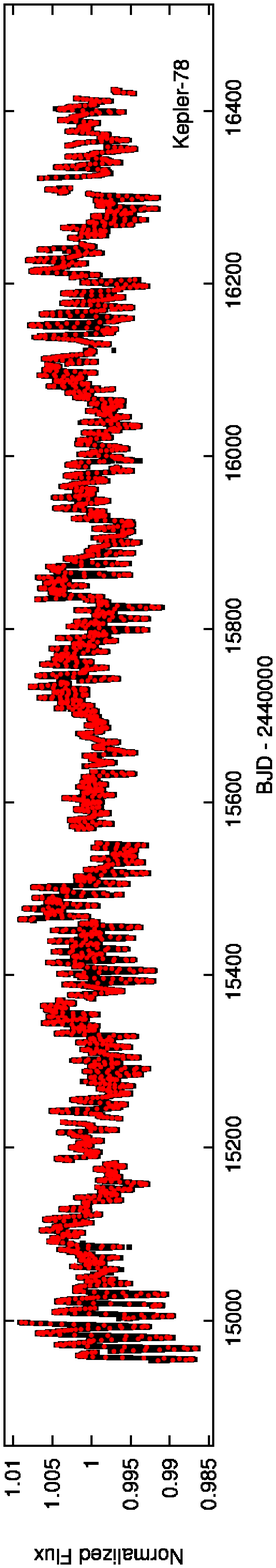}
\includegraphics[scale=0.47, angle = 270]{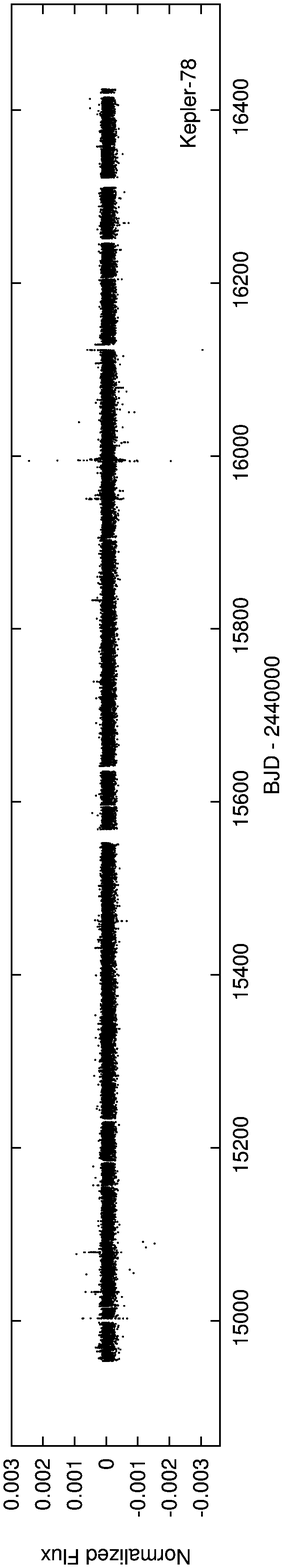}
\includegraphics[scale=0.42, angle = 270]{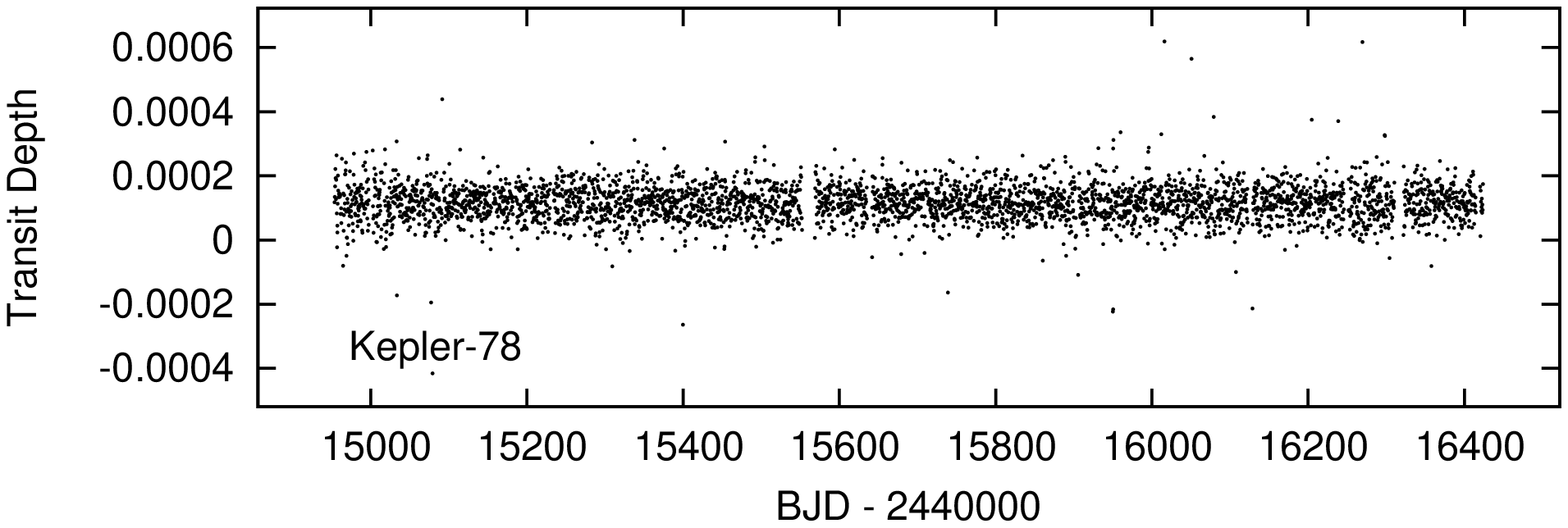}
\includegraphics[scale=0.42, angle = 270]{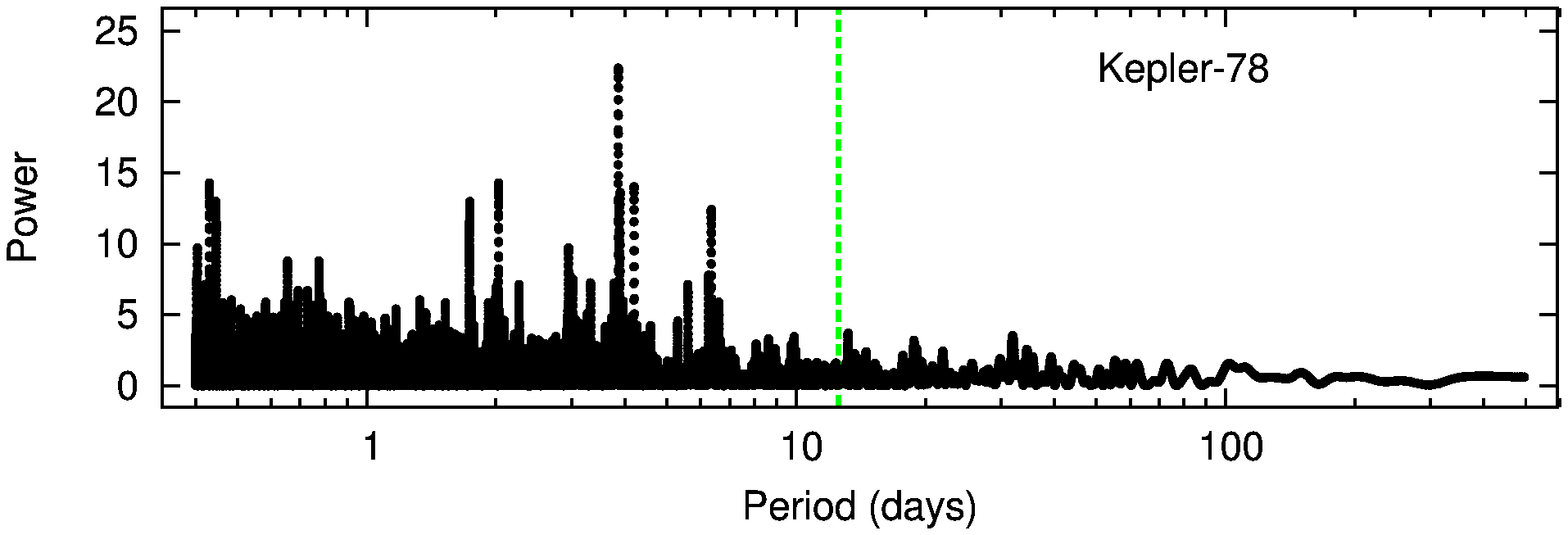}
\caption[]
	{
		The
		Kepler-78
		long cadence photometry (top), the photometry after the removal
		of the rotational modulation (middle), transit-depths (bottom left) and 
		the Lomb Scargle periodogram of the transit depths (bottom right).
		The Figure format is identical to Figure \ref{FigKIC}. 
		The vertical dotted green line in the bottom right panel
		indicates the
		$\sim$\KeplerSevenEightRotationPeriod \ d
		rotation period of Kepler-78.
 	}
\label{FigKepler78}
\end{figure*}

To determine the impact of spots on the TDRM signal of these other stellar systems
we repeat our analysis (Section \ref{SecKawahara}) for these systems; minor differences in the parameters of our analysis
to compensate for the properties of these different systems are summarized in Table \ref{TableAnalysis}.
We display our analysis results in
Figure \ref{FigKepler17} for Kepler-17,
Figure \ref{FigHATP11} for HAT-P-11,
and Figure \ref{FigKepler78} for Kepler-78.
Kepler-17 displays an obvious peak in the Lomb Scargle periodogram at the rotation period of the star.
There is a modest peak at the rotation period in the HAT-P-11 Lomb Scargle periodogram, while
for Kepler-78 no peak is apparent at the rotation period.
We note that the lack of a peak at the rotation period for Kepler-78, indicates that our technique for removing the 
rotational spot modulation
does not induce a prominent signal into our measurements of the transit depths (at least at a detectable level for shallow transits, such as 
those displayed by Kepler-78). We present further evidence that our transit depths are not affected by leakage from 
the rotational modulation below in Section \ref{SecBootstrap}.

Ergo, the KIC 1255 TDRM signal does not appear to be entirely unique to KIC 1255; 
Kepler-17 displays a prominent
correlation between the transit depths and the rotation period of the star due to occulted spots. 
This analysis indicates the possibility that occulted spots
could be the source of the peak in the Lomb Scargle periodogram of the KIC 1255b transit depths near the
stellar rotation period.

% In addition, the transit depths of exoplanets orbiting spotted stars will will be modulated by unocculted spots.

%EMULATEAPJCHANGE, 0.42 to 0.40, for the bottom ones...

\subsection{Is the KIC 1255 TDRM signal statistically significant?}
\label{SecBootstrap}

\begin{deluxetable}{cc}
\tablecaption{Fraction of Bootstrap Iterations}
\tabletypesize{\scriptsize}
\tablehead{
\colhead{Star}  & \colhead{Fraction of bootstrap signals}	\\	
\colhead{}  	& \colhead{above the original signal}		\\	
}
\startdata
KIC 1255	& 0/1000	\\
Kepler-17	& 0/1000	\\
HAT-P-11	& 49/1000	\\
Kepler-78	& 497/1000	\\
\enddata
% \tablenotetext{a}{We fix $t_{transit}$ to the predicted mid-point of the transit for this analysis, due to the fact we are unable to detect the transit on this occasion.}
\label{TableBootstrap}
\end{deluxetable}

%EMULATEAPJCHANGE, clockwise...
\begin{figure}
\centering

\includegraphics[scale=0.43, angle = 270]{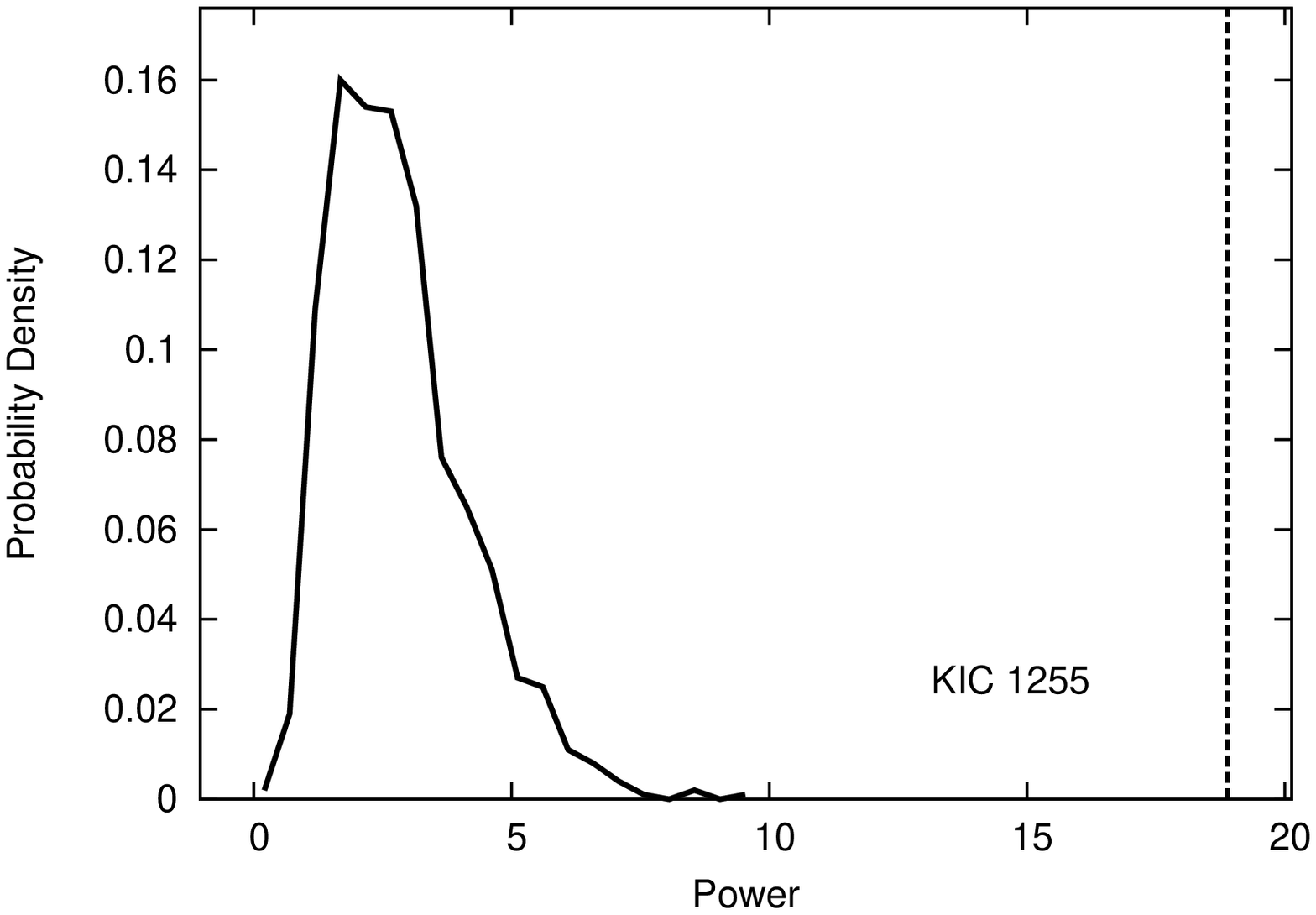}
\includegraphics[scale=0.43, angle = 270]{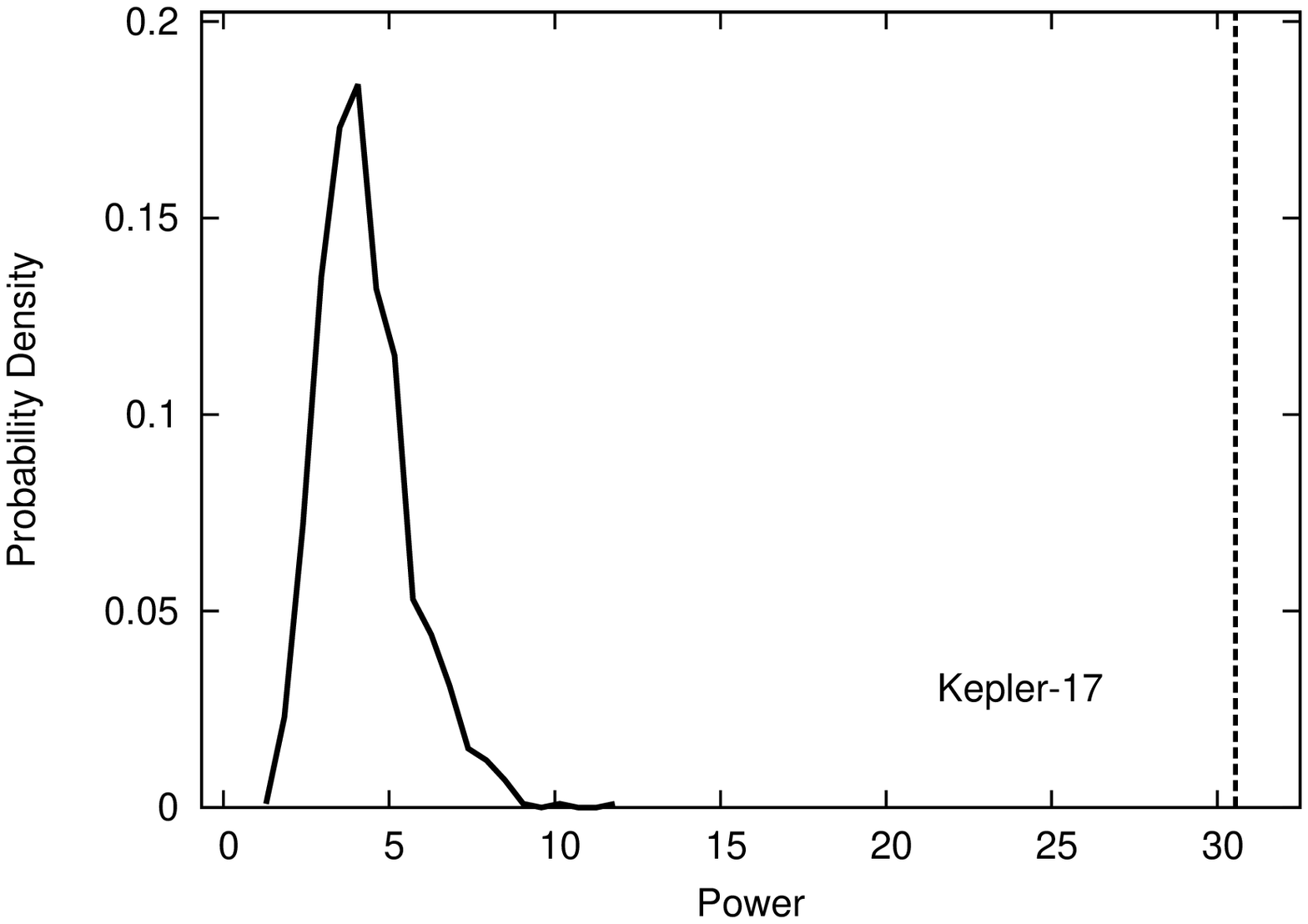}	%Kepler-17
\includegraphics[scale=0.43, angle = 270]{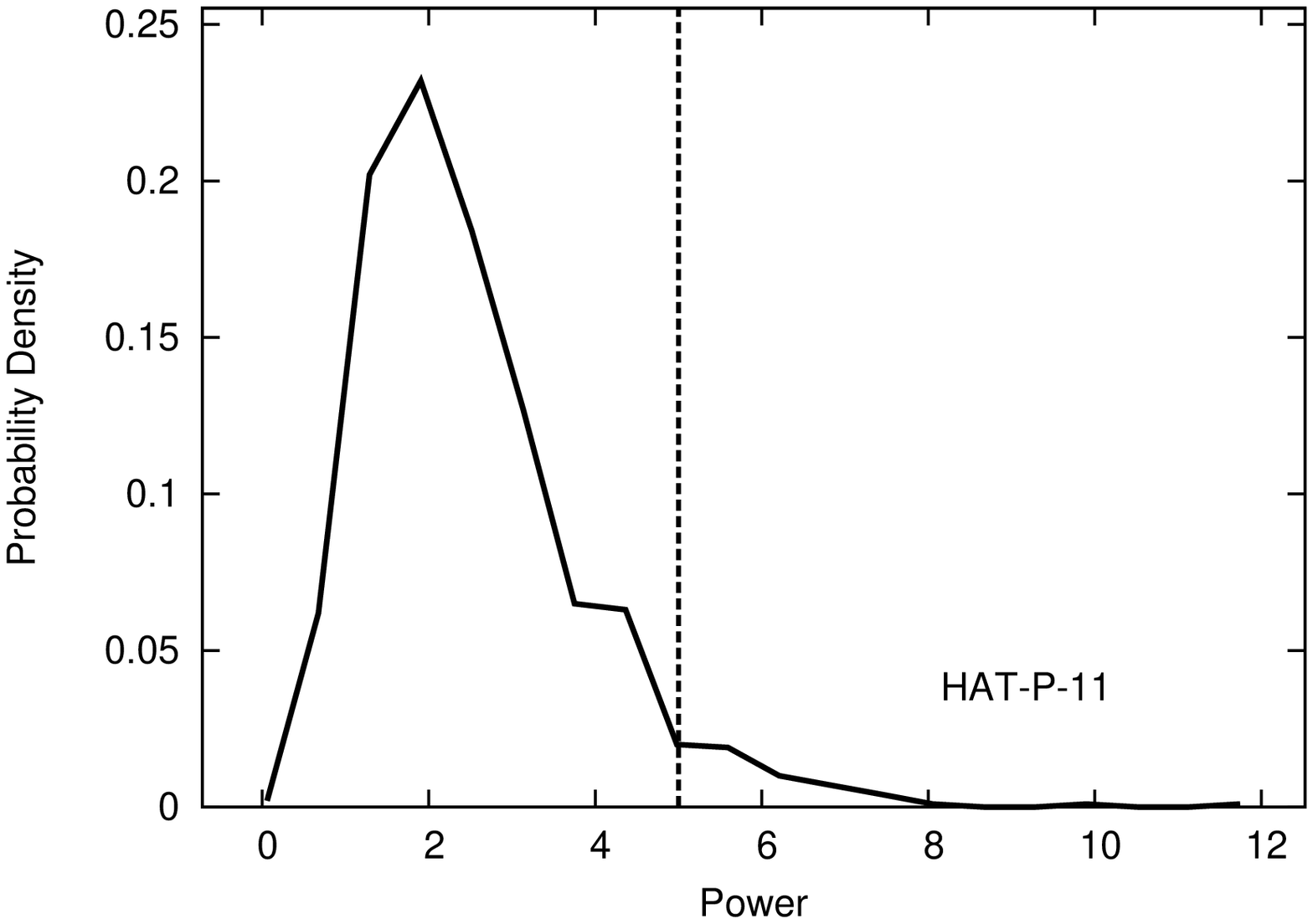}	%HAT-P-11
\includegraphics[scale=0.43, angle = 270]{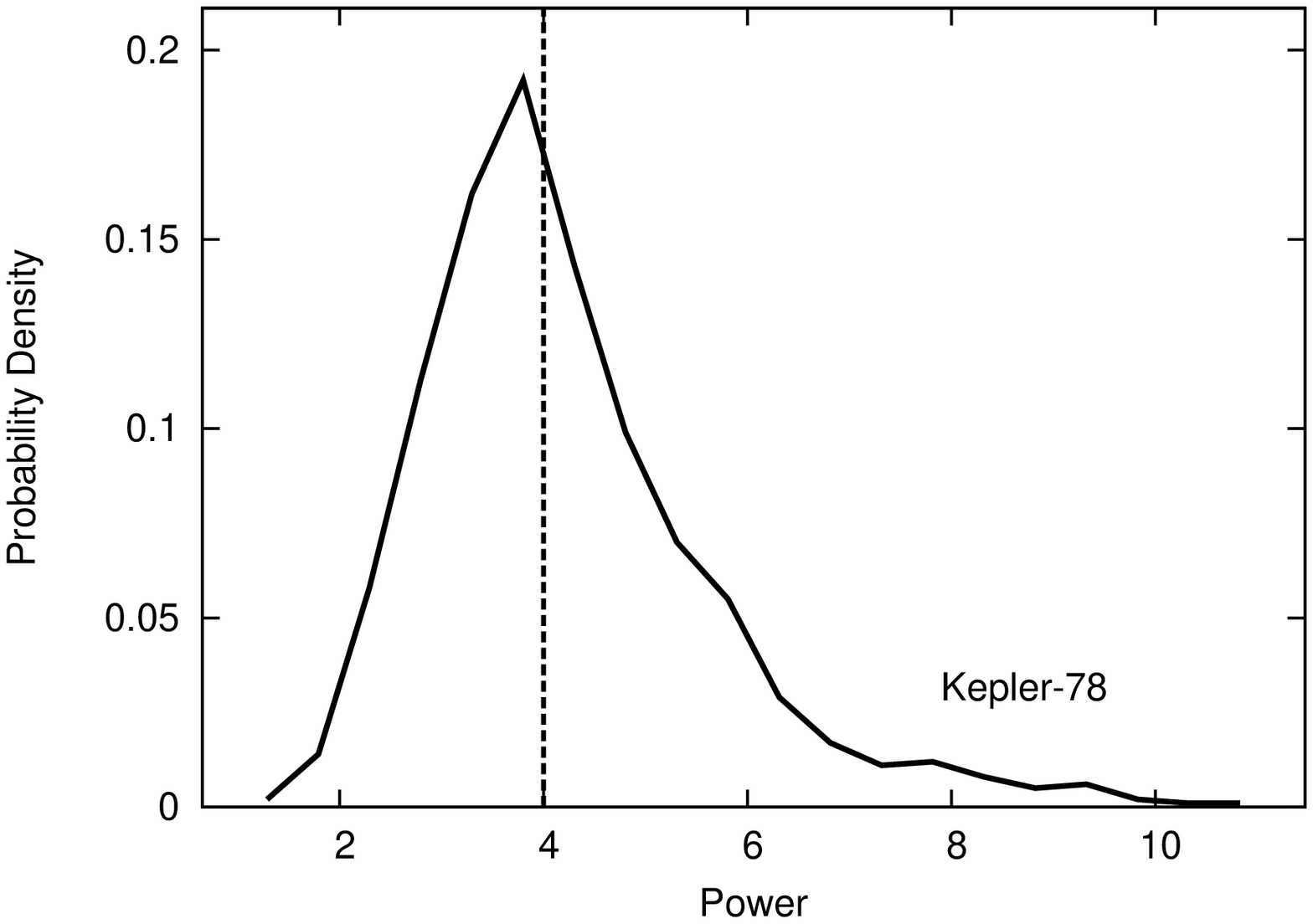}	%Kepler-78
\caption[]
	{
		A histogram of the maximum Lomb Scargle periodogram power of the transit depths near the rotation period 
		of the various stars following scrambling of the transits using bootstrap techniques.
		These stars, from top to bottom, are: KIC 1255, Kepler-17, HAT-P-11 and Kepler-78.
		%These stars, clockwise from top left, are: KIC 1255, Kepler-17, HAT-P-11 and Kepler-78.
		The vertical dashed line displays the Lomb Scargle periodogram power for the original unscrambled data.
		The transit-depth rotational modulation signal is highly significant
		for KIC 1255, but it also highly significant for Kepler-17, where the signal
		is likely due to occulted spots.
	% from a period of \KICRotationPeriodShort \ - \KICRotationPeriodLong \ d 
 	}
\label{FigHisto}
\end{figure}

 To determine if the TDRM signal is statistically significant,
we repeat the analysis we present in Section \ref{SecKawahara}, but 
randomly scramble and interchange the transits for one another.
That is, after removing the rotational modulation using our cubic spline,
we randomly scramble the data from a given transit for another using bootstrap techniques,
reintroduce the rotational modulation, and then subject the data to our Section \ref{SecKawahara} analysis.
To interchange the data during transits,
once the rotational modulation has been removed using our cubic spline fit,
we trade the data between the phases of $\phi$ = 0.4 - 0.7 of one transit with another transit
that is chosen at random.
The only obvious drawback of this technique is that the bootstrap data will no longer have the same strict 
$\sim$29.4 minute sampling at the beginning and end of the transit
as the original KIC 1255 {\it Kepler} long cadence data.
Nonetheless, we then compare
the strength of the largest peak in the Lomb Scargle periodogram near the rotation period of the scrambled transits
to that of the original unscrambled transits.
We accept values for the peak of the Lomb Scargle periodogram within a day of the rotation period of the star, so we take the 
maximum Lomb Scargle periodogram
power from periods of \KICRotationPeriodShort \ - \KICRotationPeriodLong \ d, compared to the 
$\sim$\KICRotationPeriod \ d
rotation period of the star.
Using these techniques we can determine what fraction of
the Lomb Scargle periodograms of the transit depths of KIC 1255b
display a signal as large as what is observed simply by chance.
We perform this bootstrap test on our KIC 1255b transits, excluding 
the data before and after the quiescent periods
as in Figure \ref{FigKICExcludeShallow}; the results are similar
if we include all the KIC 1255b transits, as in Figure \ref{FigKIC}.

A histogram of the maximum Lomb Scargle periodogram power near the KIC 1255 stellar rotation period
for 1000 bootstrap iterations of our scrambled
KIC 1255b transit depths is displayed in Figure \ref{FigHisto}.
Zero of \BootstrapScrambleIterations \ bootstrap iterations %\BootstrapScramble \
achieve a power as high as the original data.
Therefore the KIC 1255 TDRM signal is highly significant.
However, this highly significant signal could still be due to occulted spots.

To demonstrate this possibility, we repeat this analysis for Kepler-78, HAT-P-11, and Kepler-17.
We accept values for the peak in the Lomb Scargle periodogram within a day of the rotation period for these stars.
We summarize the results in 
Table \ref{TableBootstrap}, and display
the results in Figure \ref{FigHisto}.
Kepler-17 also displays a highly significant TDRM signal that, as previously discussed in Section \ref{SecStars},
is likely simply due to occulted spots. The HAT-P-11 TDRM signal, again likely due to occulted spots,
is detected at nearly the 2$\sigma$ level.

We note that as we reintroduce the rotational modulation after we have scrambled the transits
in the bootstrap data-sets submitted to our analysis,
the low power observed
for many of the scrambled transit bootstrap datasets 
indicates that our analysis is efficient at effectively removing the impact
of the rotational modulation on our measured transits depths. This further reiterates the point
that the observed signal is not due to an artefact of 
leakage from the observed rotational modulation into our measured
transit depths.

% \citet{Kawahara13} presented simple arguments that the observed 30\% variation in the transit depths over
% the rotation period of the star was much larger than the expected 4\% effect due to unocculted starspots. 

\subsection{Transit-timing analysis}
\label{SecTiming}

%EMULATEAPJCHANGE, 0.42 to 0.40
\begin{figure*}
\centering
\includegraphics[scale=0.42, angle = 270]{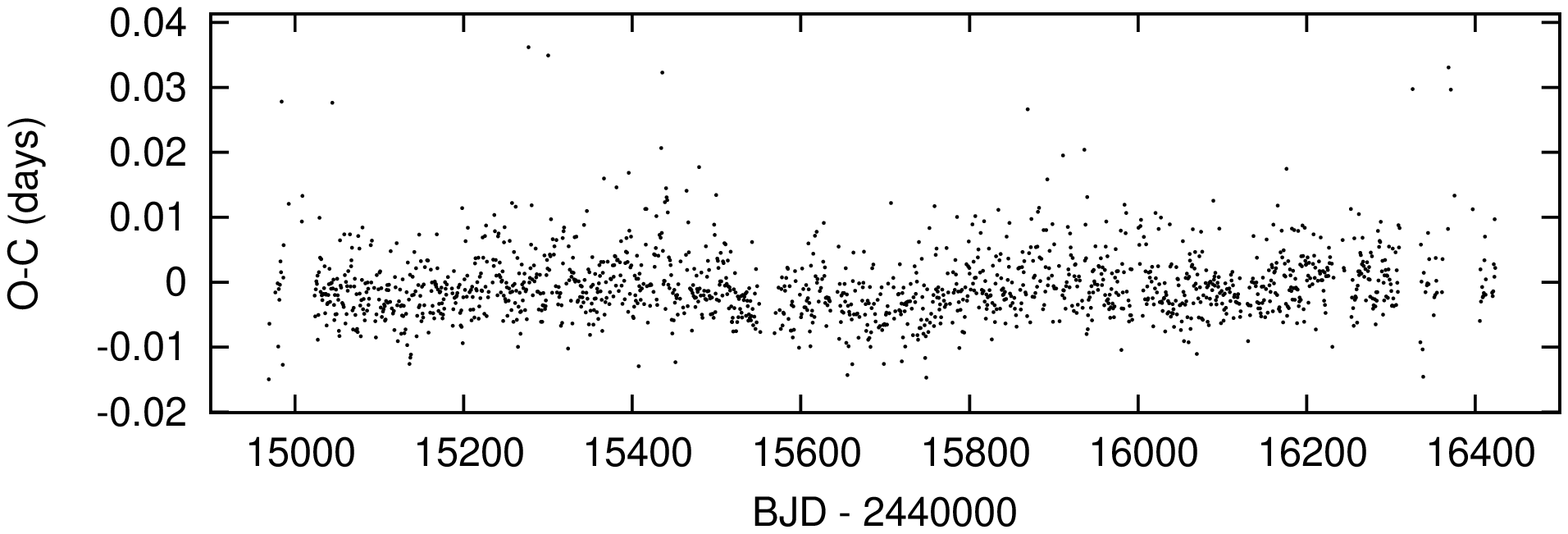}
\includegraphics[scale=0.42, angle = 270]{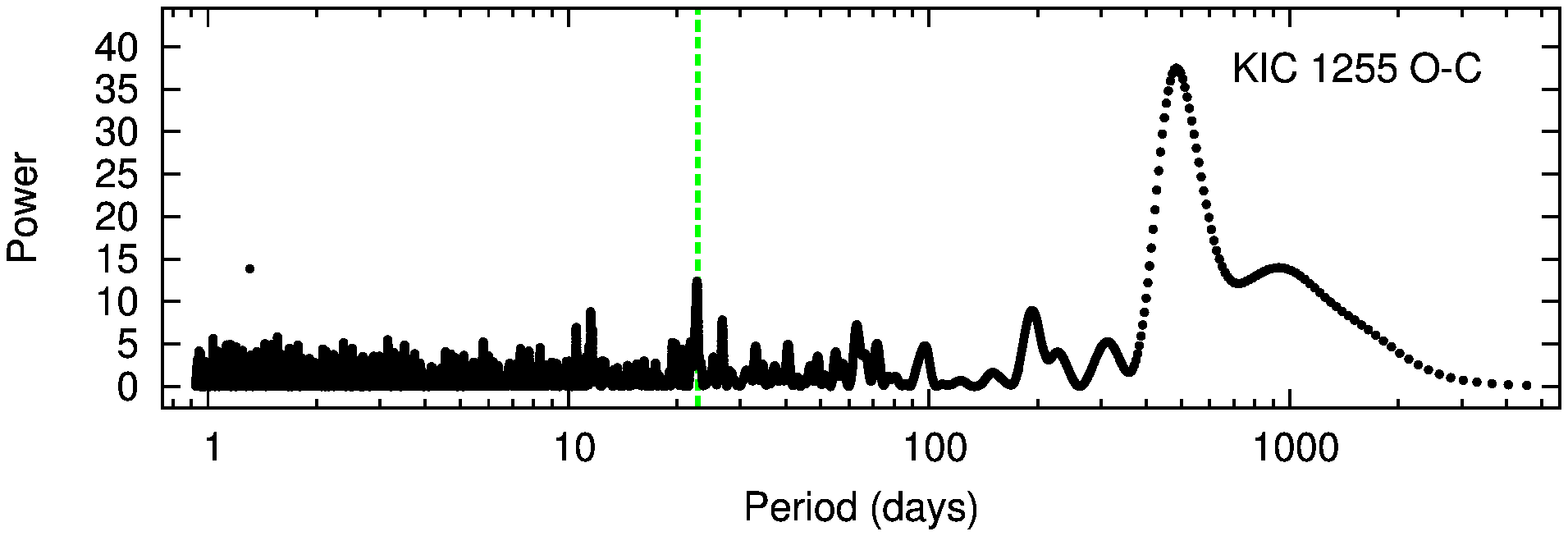}
\includegraphics[scale=0.42, angle = 270]{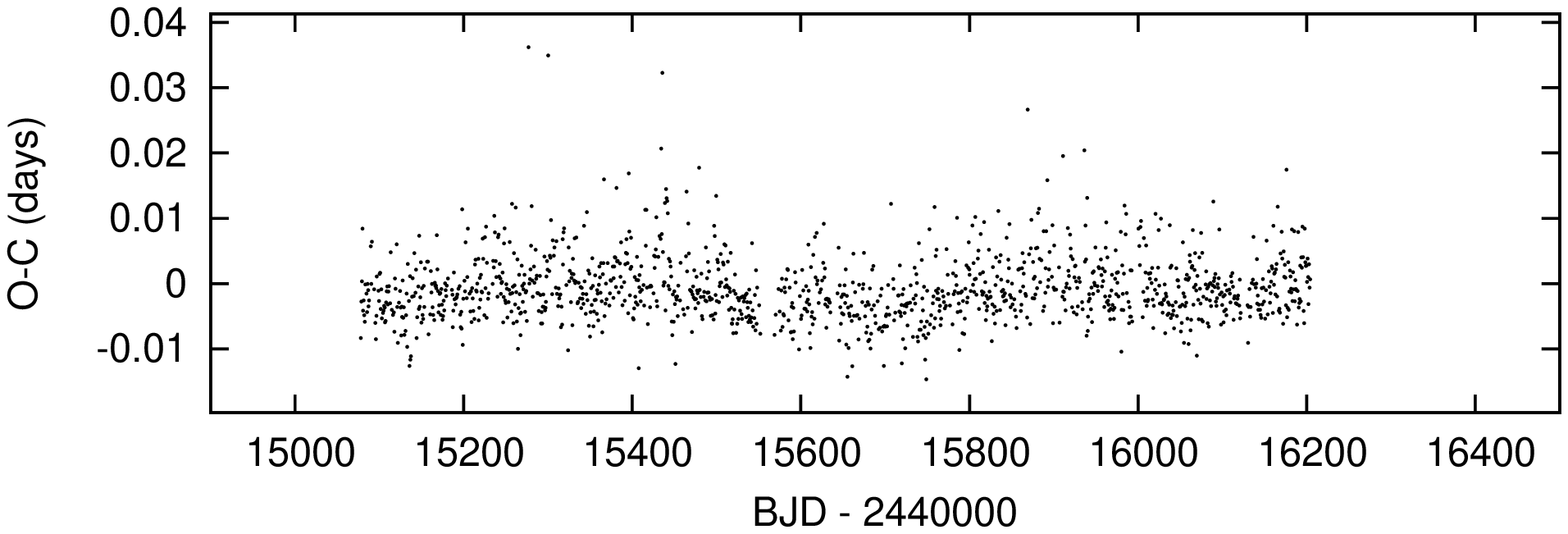}
\includegraphics[scale=0.42, angle = 270]{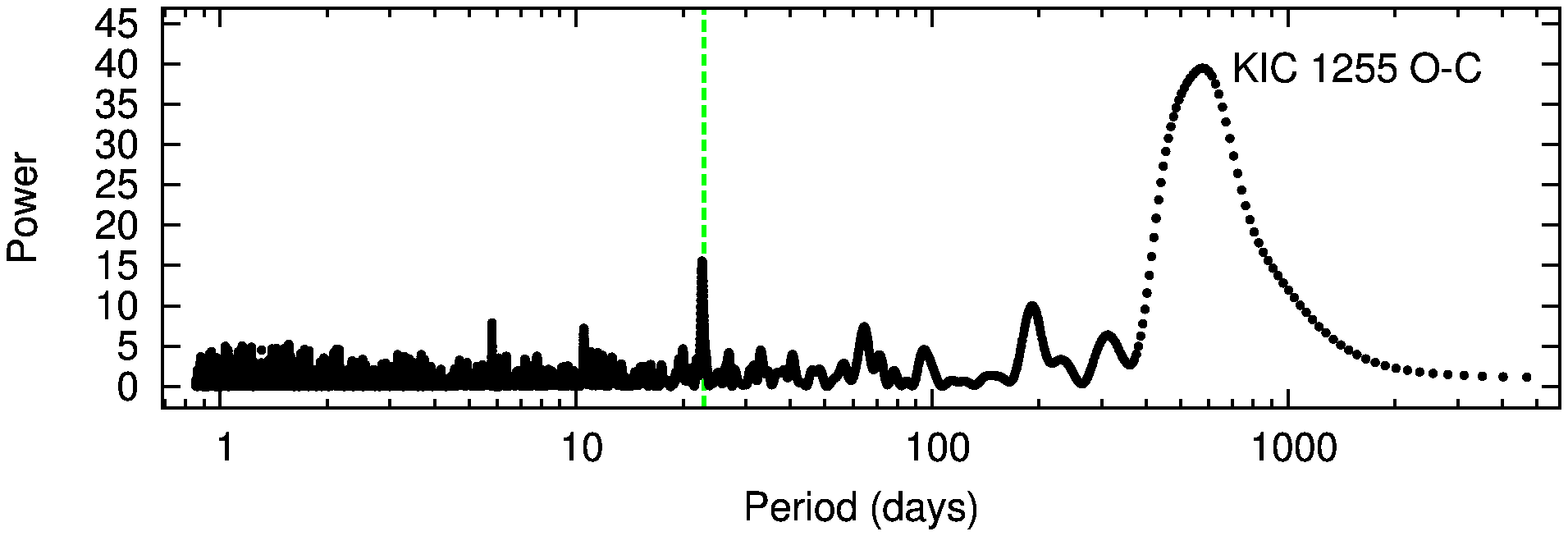}
\includegraphics[scale=0.42, angle = 270]{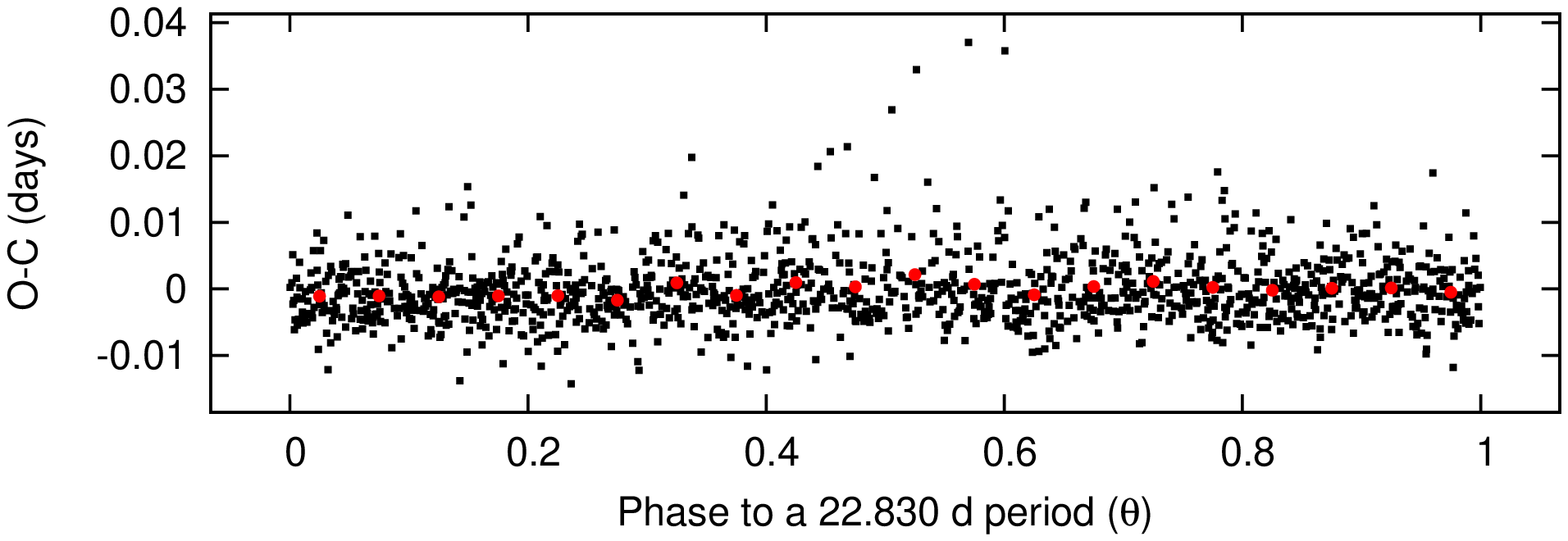}
\includegraphics[scale=0.42, angle = 270]{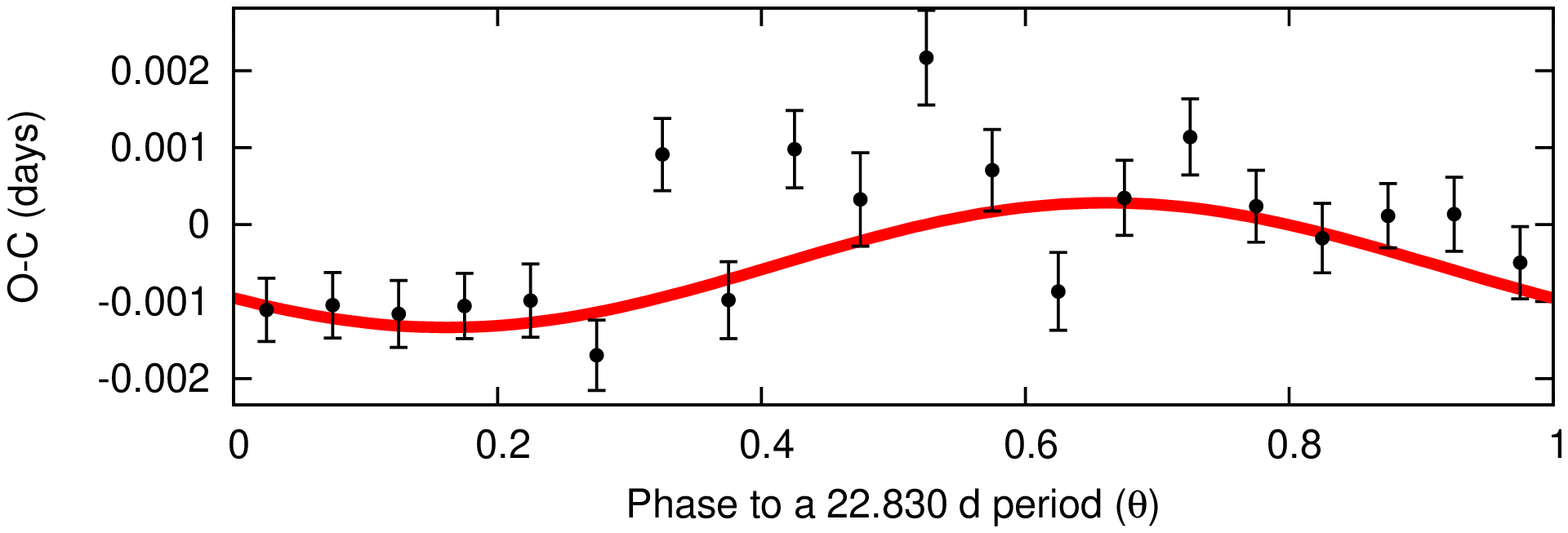}
\caption[]
	{
		Top left panel: the transit timing analysis of all the transits of KIC 1255b.
		Top right panel: a Lomb Scargle periodogram of the O-C timing residuals
		of the transit-times of KIC 1255b. The vertical dotted green line denotes
		the rotation period of KIC 1255. 
		Middle panels: the transit timing analysis for KIC 1255b excluding the quiescent periods (left panel),
		and the associated Lomb Scargle periodogram (right panel).
		Bottom left panel: the O-C transit times of KIC 1255b, excluding the quiescent periods,
		phased to the rotation
		period of the star (black points); the red points display the data binned every
		0.05 in phase. 
		Bottom right panel: The same binned data every 0.05 in phase, with the 3$\sigma$ upper limit
		on the best-fit sine-curve displayed with a red solid line.
 	}
\label{FigTiming}
\end{figure*}

We also perform a timing analysis of all the {\it Kepler} long cadence transits of KIC 1255b (Figure \ref{FigKIC}),
and the {\it Kepler} transits excluding the quiescent
periods (Figure \ref{FigKICExcludeShallow}).
After removing the rotational
modulation of KIC 1255 with our cubic spline as described in Section \ref{SecKawahara}, we fit 
all the long cadence {\it Kepler} photometry of KIC 1255 with a scaled and shifted version
of the mean long cadence {\it Kepler} profile of the transit of KIC 1255b (as displayed in Figure 
1 of \citealt{CrollKIC}).
We fit to
Equation 1 of \citet{CrollKIC}
using Markov Chain Monte Carlo (MCMC) techniques as described in \citet{CrollMCMC}.
We use MCMC chains with 150000 steps, which we found to be sufficient to return properly determined transit-times
and errors.
We place an a priori constraint on the offset from the expected 
mid-point of each transit of KIC 1255b of $\pm$ 0.04 d (58 minutes).
We cut out all transits where the error on the timing of the transits exceeds 0.008 d ($\sim$11.5 minutes),
%leaving 1317 transit-times. %Short cadence profile
leaving 1571 transit-times. %Long cadence profile, all the Kepler data...
This cut generally serves to cut out shallow transits, where it is difficult to identify the transit,
and therefore challenging to return a useful limit on the timing of the transit mid-point.
% leaving 1029 transits of the original list of 1604 transits.
%, where $A$ denotes the scaled depth of the transit, normalized to the mean transit depth
%of $A$=1, which corresponds to a transit depth of approximately $\sim$0.55\% of the normalized flux of the star.
%As the timing of transits with depths below $A$=0.5 (or 0.28\% of the normalized flux) are poorly 
%constrained, we remove
%these transits from our transit-timing analysis.

 The transit-timing results are displayed in the top and middle left panels of 
Figure \ref{FigTiming} for all the transits, and for the transits excluding the quiescent periods.
We also perform a Lomb 
Scargle periodogram on the transit-timing results to search for periodicities in the transit-timing
signal. The results
are shown in the top and middle right panels of Figure \ref{FigTiming}.
The most prominent signal is at $P$$\sim$563 d.
There is a modest signal at the rotation
period of the star; we fit a sinusoid to this signal and place a 3$\sigma$
upper-limit on its amplitude of $\sim$\TransitTimingLimitSeconds \ s, after scaling
the errors\footnote{We increased our error bars by a factor of $\sim$2.5.}
so the reduced $\chi^2$ is equal to 1.0
(the bottom panels of Figure \ref{FigTiming}).
If we perform the same analysis on only the first half of the data
(excluding the quiescent periods; BJD - 2440000 = 15078 - 15600),
the associated
3$\sigma$ limit on the amplitude is $\sim$\TransitTimingLimitFirstHalfSeconds \ s.

\subsubsection{The orbital period and a limit on orbital decay}

Our transit-timing analysis also allows us to measure the best-fit orbital period
of KIC 1255b, $P$, and place a limit on the change of the period with time, $\dot{P}$, by
fitting linear and quadratic functions to the integer transit number and the dates of the transit mid-points.
The best-fit period is: 
%$P$ = 0.653,553,2(2) d, %Short cadence profile - fit correctly
%$P$ = 0.653,552,9(2) d, %Long cadence profile - fit correctly, dataset_want_int =0
$P$ = 0.653,553,4(2) d, %Long cadence profile - fit correctly...
and the associated mid-point of the transit is BJD=2454968.9820(7)\footnote{The mid-point of the transit is defined
as the minimum flux point of the mean {\it Kepler} Long cadence transit profile (Figure 1 of \citealt{CrollKIC}).},
where the value in brackets in both cases is the 1$\sigma$ error on the last digit.
Our best-fit value on the change in period with time divided by the period is:
% $\dot{P}$/$P$ = 2.4 $\pm$ 0.8 $\times$$10^{-9}$ $d^{-1}$. %ShortCadence
%$\dot{P}$/$P$ = 2.6 $\pm$ 1.3 $\times$$10^{-9}$ $d^{-1}$. %LongCadence 
%$\dot{P}$/$P$ = 2.8 $\pm$ 0.8 $\times$$10^{-9}$ $d^{-1}$. %LongCadence dataset = 5
$\dot{P}$/$P$ = 2.4 $\pm$ 0.7 $\times$$10^{-9}$ $d^{-1}$, %LongCadence dataset = 5, 2c
suggesting that, if anything, the period of the planet is increasing; 
therefore, there is no evidence that the orbit of the planet is rapidly decaying.

% The associated 3$\sigma$ lower-limit on $P$/$\dot{P}$ is then
% % 6$\times$$10^{5}$ years, %ShortCadence
% %4$\times$$10^{5}$ years, %LongCadence
% 6$\times$$10^{5}$ years, %LongCadence dataset = 5
% suggesting that the orbit of the planet is not rapidly decaying.

\subsubsection{Profile of Early, Normal and Late Transits of KIC 1255b}

\begin{figure}
\centering
\includegraphics[scale=0.42, angle = 270]{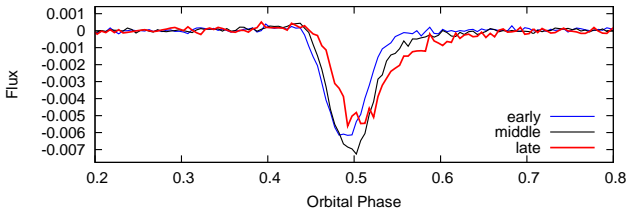}
\caption[]
	{
		The long cadence {\it Kepler} photometry of KIC 1255 phased 
		to the orbital period of the planet.
		The mean transit profile of the ``early'' transits ($O-C$ $<$ -0.003 d) are shown with the blue line,
		the ``middle'' transits (-0.003 $<$ $O-C$ $<$ 0.003 d) are shown with the black line,
		and the ``late transits ($O-C$ $>$ 0.003 d) are shown with the red line.
 	}
\label{FigEarlyMiddleLate}
\end{figure}

In order to determine if there is an astrophysical reason for the transit-timing variations, we
also present the mean {\it Kepler} long cadence transit profile for those transits
that occur earlier than expected, as expected, and later than expected.
We present the
``early'' ($O-C$ $<$-0.003 d; 418 transits),
``middle'' (-0.003 $<$ $O-C$ $<$ 0.003 d; 832 transits) and
``late'' ($O-C$ $>$ 0.003 d; 321 transits) transit profiles of KIC 1255b in Figure \ref{FigEarlyMiddleLate}.
The fact that the ingress of the ``early'' and ``middle'' transit profiles occur at approximately the same time is consistent
with the candidate planet hypothesis with a tail of dust streaming behind it; in this scenario the ingress of the transit
should occur no earlier than the ingress of the Mercury-sized planet across the stellar disk.
The ``late'' transit profile could be due to an 
brief ($\sim$hour long)
cessation of dust emission before the start of the ingress, leading to the dust tail trailing further behind
KIC 1255b than usual\footnote{On a more
speculative note, the longer egress tail in the ``late'' transit profile could imply a
higher than normal dust ejection rate for several hours prior to the transit
followed by a period of exhaustion of the dust supply; similarly \citet{vanWerkhoven14} noted
that there were a number of occasions in the {\it Kepler} data
of ``on-off' transit behaviour, 
where relatively deep transits of KIC 1255b were followed by undetectable transits.}.
% On a more
% speculative note, the longer egress tail in the `late' profile could imply a
% higher than normal dust ejection rate for several hours prior to the transit
% followed by a period of exhaustion of the dust supply (to explain the late
% start).

% Despite the fact the transit-times for these three profiles 
% are very different by design, the very start of the ingress of the transit appears to occur at the same
% time for each of the transit profiles.

\section{Possible Explanations for the KIC 1255 Transit Depth Rotational Modulation Signal}
\label{SecExplain}

\subsection{Could the TDRM signal be due to unocculted spots?}
\label{SecUnocculted}

The flux from a spotted star with a transiting planet, $F(t)$, at time $t$ can be
written as \citep{Carter11}:
\begin{equation}
F(t) = F_o [1 - \epsilon(t)] - \Delta F(t)
\end{equation}
where $F_o$ is the flux of the unspotted and untransited star,
$\epsilon (t)$ is the fractional loss of light due to starspots, and $\Delta F(t)$
is the flux blocked (or scattered in our case) by the planet during transit.
As we normalize our signal by subtracting the mean of the long cadence photometry 
using our cubic spline fit at a given time, $t$,
we essentially\footnote{As Figure \ref{FigSpotsLightcurve} indicates, KIC 1255 is a very spotted star, and it
is possible that there are always spots visible on KIC 1255 during the {\it Kepler} light curve.
If this is the case, we will have underestimated $\epsilon(t)$; in
that case our estimate of the transit depths of KIC 1255b will be systematically overestimated, but these
depths will not
be modulated at the rotation period, so this bias would not affect the rest of our conclusions.}
subtract $\epsilon(t)$, and our measurement of the transit depths should
be unaffected by unocculted spots\footnote{There will be a modest effect due to the normalization
of the {\it Kepler} light curve by dividing through each {\it Kepler} quarter by its mean flux level. 
In this case, we would expect the transit depths to be modulated by at most a factor of $\frac{1}{1-\epsilon(t)}$, which
for $\epsilon(t)$ $\sim$ 1-5\% is much less than the TDRM signal \citep{Kawahara13}.}.
The TDRM signal should not be due to unocculted spots.

% As KIC 1255 displays significant rotational modulation, one would usually expect
% that the transit depths of KIC 1255
% would be modulated at the rotation period of the star, simply due to unocculted spots.

\subsection{Could the TDRM signal be due to occulted spots?}
\label{SecOcculted}

% \begin{figure}
% \centering
% \includegraphics[scale=0.50, angle = 270]{kawahara_bootstrap_histogram_datasetwant_0_bootstrap_typecare_1.eps}
% \caption[]
% 	{
% 		A histogram of the maximum Lomb periodogram power of the KIC 1255b transit depths
% 		near the stellar rotation period 
% 		following modulating the transit depths simulating the effects
% 		of occulted spots.
% 		The vertical dashed line displays the Lomb periodogram power for the original KIC 1255 data.
% 		Occulted spots can possibly account for the KIC 1255 TDRM signal.
% 	% from a period of \KICRotationPeriodShort \ - \KICRotationPeriodLong \ d 
%  	}
% \label{FigHistoOccult}
% \end{figure}

Although the signal of a transiting planet occulting a cool starspot
often leads to significant deviations (brightenings) in the transit light curve, these deviations
are often short compared to the duration of the transit. However, due to the long cometary tail supposedly trailing behind
the candidate planet KIC 1255b, if a spot occultation occurs, it should cause an anomalous brightening
for a significant fraction of the transit
duration. In this way, one might expect that occulted spots would play a larger role in modulating
the transit depths of KIC 1255b at the stellar rotation period
than would normally be expected for a transiting planet.

The fractional brightening of the normalized light curve
during a transit due to occulted spots can be described as:
\begin{equation}
\delta F/F_* = \frac{A_{spot} (t)}{\pi R_*^2} \times (1 - \frac{I_s}{I_*})
\end{equation}
where $A_{spot} (t)$ is the area of the spot projected onto the viewing plane
that is occulted by the planet at time, $t$,
$R_*$ is the radius of the star, and 
$I_s$ and $I_*$ are the intensity of the spot and star, respectively.
The diffuse cloud believed to be trailing the candidate planet KIC 1255b will not block light
from the star per se, but the scattering from the dust tail has a similar effect.

For the sake of simplicity, let us treat
the cometary tail as a thin rectangular ribbon,
and let us assume
that the planet and its cometary tail are completely opaque,
 and the planet has a nominal transit depth of $\Delta F(t)$.
Let $\Delta F(t)$ be measured at the
point when the cometary tail stretches completely
across the star, and thus blocks out a width, $w$, on the star of $2 R_*$ and a height, $h$, where $h$ $<<$ $R_*$.
In that case the planet and tail cover a fractional area of the star, equivalent to the transit depth, $\Delta F(t)$, of:
\begin{equation}
\Delta F(t) = \frac{h \times w}{\pi R_*^2} = \frac{ 2 h}{\pi R_*}
\end{equation}
so, $h$ = $\pi$ $\Delta F(t)$  $R_*$/2. In the \citet{Rappaport12} cometary tail model of KIC 1255b,
the scattering material is not believed to be completely opaque, but in that case this will just serve to enlarge $h$,
and the effect
will be qualitatively equivalent.
Compared to the case where no spot is occulted, the expected increase 
in flux during transit due to the fraction of the spot that is occulted, of area $A_{spot} = w_s \times h \sim 2 R_s h$, when $h$ $<$ $R_s$,
is then\footnote{If we instead assume a dust tail with an exponential falloff
in density (with scale length $\ell$), then Equation \ref{EquationOcculted} would simply be
multiplied by a factor of $2 \xi e^{-f \xi} (1-e^{-2 \xi})^{-1}$ where $\xi
\equiv R_*/\ell$, $f \equiv L/R_*$, and $L$ is the distance between the
planet and the starspot.  This factor ranges between $\sim$0.2 and 2 for
plausible choices of $L$ and $\ell$, with a most likely value of $\sim$1.}:
\begin{eqnarray}
\delta F/F_* \sim \frac{ 2 R_s h}{\pi R_*^2} \times (1 - \frac{I_s}{I_*}) \nonumber \\
\sim \frac{2 R_s \times \Delta F(t) \times \pi R_* /2}{\pi R_*^2} \times (1 - \frac{I_s}{I_*}) \nonumber \\
\sim \ \Delta F(t) \frac{R_s}{R_*} \times (1 - \frac{I_s}{I_*})
\label{EquationOcculted}
\end{eqnarray}
If we assume a single spot, then the size of the spot is proportional to the maximum fractional loss of light
due to starspots for that rotational cycle, $\epsilon$:
\begin{equation}
\epsilon \sim \frac{R_s^2}{R_*^2} \times (1 - \frac{I_s}{I_*})
\end{equation}
and Equation (\ref{EquationOcculted}) reduces to:
\begin{equation}
\delta F/F_* \sim \Delta F(t) \sqrt{\epsilon (1 - \frac{I_s}{I_*})}
\end{equation}
The rotational modulation that we observe is $\sim$1-5\%;
$\epsilon$ might actually be larger than this, as Figure \ref{FigSpotsLightcurve}
suggests a multitude of spots could be present on KIC 1255, and masking the true unspotted flux
level of the star.

The most extreme assumption would be spots that are completely dark: $I_s$ = 0.
If we use this assumption and use an estimate of $\epsilon$$\sim$4\%, this results in spots of size: $R_s$ $\sim$ 0.2 $R_{*}$.
Using $I_s$ = 0 and $\epsilon$ $\sim$ 4\%, Equation (\ref{EquationOcculted}) becomes simply:
$\delta F/F_*$ $\sim$ $0.2 \Delta F(t)$.
That is the occulted spot signal, at its maximum, could represent 
a 20\% brightening compared to the depth of the transit; the TDRM
signal is 25\% of the depth of the transit.
Occulting a cool spot that modulates the flux of KIC 1255 at approximately the 4-6\% 
level is therefore likely sufficient to explain the TDRM signal.

Occulted spots are known to cause modest transit-timing variations,
generally on the order of a minute or less for conventional transiting
planets \citep{Barros13}; however, due to KIC 1255b's unique geometry
and the putative cometary tail trailing behind it, we decided to explicitly
model the effect of occulted spots on the transit light curve.
We considered an occulted spot of radius in the range of $R_{s} =
0.20-0.25 \, R_*$ with an internal surface brightness of 20\% that
of the host star ($I_s/I_*$=0.2).
We then simulate a comet-like tail that is more narrow in
height than $R_{*}$ and allow it to pass over a limb-darkened\footnote{Using a linear limb-darkened profile
with a limb-darkening coefficient of $u$=0.8.}
stellar disk at the
same latitude as the spot.  We adopt an exponentially decreasing dust
density in the tail with increasing distance from the planet.  The transit
profile is then calculated for the chosen spot longitude; after which the
profile is convolved with the {\em Kepler} long cadence integration time.  Finally,
the spot is stepped systematically in longitude and the resultant transit
profiles are recalculated. We then subject these simulated profiles to a
transit-timing analysis similar to what is used for the actual transits.

Our estimate suggests that occultations of a single spot causing 4\% rotational modulation, results in
approximately 25\% variations in the transit depth, and transit-timing
variations of up to $\sim$\TransitTimingVariationsOccultedSpotsSeconds \ s.  Transit-timing variations this large
are ruled out by our 3$\sigma$ limit of \TransitTimingLimitTwiceSeconds \ s
on the peak-to-peak timing variations on the transit of KIC 1255b phased to the
stellar rotation period (Figure \ref{FigTiming}) for the entire data-set. Such large
transit-timing variations are not ruled out, however, if we only analyze the first-half of the
{\it Kepler} long cadence data excluding the quiescent periods (BJD -
2440000 = 15078 - 15600), where the TDRM signal is stronger (Figure \ref{FigKawaharaSplit}).
Also, if multiple spots are occulted then transit depth variations
of 25\% can easily be caused without detectable transit timing variations.

% adjusted the mean
% transit profile of KIC 1255b (Figure 1 of \citealt{CrollKIC}) using Equation \ref{EquationOcculted}
% to simulate the effects of occulted spots on the transit profile and then submitted
% these data to the transit fitting analysis we discuss in Section \ref{SecTiming}.
% A spot was assumed to be placed at several different longitudes on the stellar surface (corresponding
% to being occulted immediately or near the end of the transit of KIC 1255b), and the planet was assumed
% to pass over a spot 0.2 $R_*$ in size in $\sim$7.5 minutes (a value reached using the orbital velocity
% of KIC 1255b assuming a circular orbit).
% This estimate suggests that spot occultations that cause approximately 20\% variations in the transit
% depth will cause transit-timing variations of up to $\sim$80 s. Transit-timing variations this large
% are not ruled out by our 
% 3$\sigma$ limit of \TransitTimingLimitTwiceSeconds \ $s$ on the 
% peak-to-peak timing variations on the transit of KIC 1255 phased to the stellar rotation period 
% (Figure \ref{FigTiming}).

\subsubsection{Occulted Spots and the Phase of the TDRM Signal}

The theory that the TDRM signal results from occulted spots is not immediately suggested
by phasing the transit depths with the observed rotational modulation (Figure \ref{FigKICPhase}).
If the TDRM signal results from occulted spots, one would naively expect that the shallowest transits would be found
when the spots are most apparent; that is, since the minimum of the flux from the observed rotational modulation should
coincide with the time the spots are most visible, this should be when there is the highest likelihood of occulted spots. 
Figure \ref{FigKICPhase} indicates that the shallowest transit depths are not observed
near the average minimum of the observed stellar flux.
However, as the transit of KIC 1255b likely only occults
a small fraction of the surface of the star, it is very possible that the planet does not occult the largest spots
visible on the star. As can be seen in Figure \ref{FigSpotsLightcurve}, spots appear to be visible at all
stellar rotation phases, and therefore occulted spots that lead to shallower transit depths
could occur at any stellar rotation phase. Ergo, 
a lack of correlation between the phase
of the largest visible spots and the shallowest transit depths of KIC 1255b is not a good
reason to reject the possibility that occulted spots might
be causing the TDRM signal.

Figure \ref{FigKawaharaSplit} indicates that the TDRM signal is significantly stronger
in the subset of the data from BJD - 2440000 = 15078 - 15600.
The shallowest transit depths are observed at a phase compared to the stellar rotation period, $\theta$, of approximately 
$\theta$ $\sim$ 0.2. This coincides with a prominent spot that is apparent
at that stellar phase that causes modulation of the light curve at approximately
the 1\% level. A spot causing a modulation of $\epsilon$ $\sim$ 1\% would result in transit depth variations
of approximately $\delta F/F_*$ $\sim$ $0.1 \Delta F(t)$ (assuming $I_s$=0) - likely too small to explain the observed signal.
However, as Figure \ref{FigSpotsLightcurve} seems to suggest that KIC 1255 could be a very spotted star, it is possible that
there are always a number of spots visible on KIC 1255 during the {\it Kepler} light curve. In that case, what we observe
as an $\sim$1\% decrement in flux due to rotational modulation, could actually be much larger; if a spot
that actually represents approximately a 4\% decrement in flux was occulted, this would be sufficiently large to explain
the TDRM signal.

\subsubsection{A Starspot model of a subset of the KIC 1255 photometry}

%EmulateApjChange 0.045
%theta = (15413.6 - 15410.6537769)/22.86 = 0.128
%Use this website to conver to .eps
%http://image.online-convert.com/convert-to-eps
\begin{figure}
\centering
\includegraphics[scale = 0.045, angle = 0]{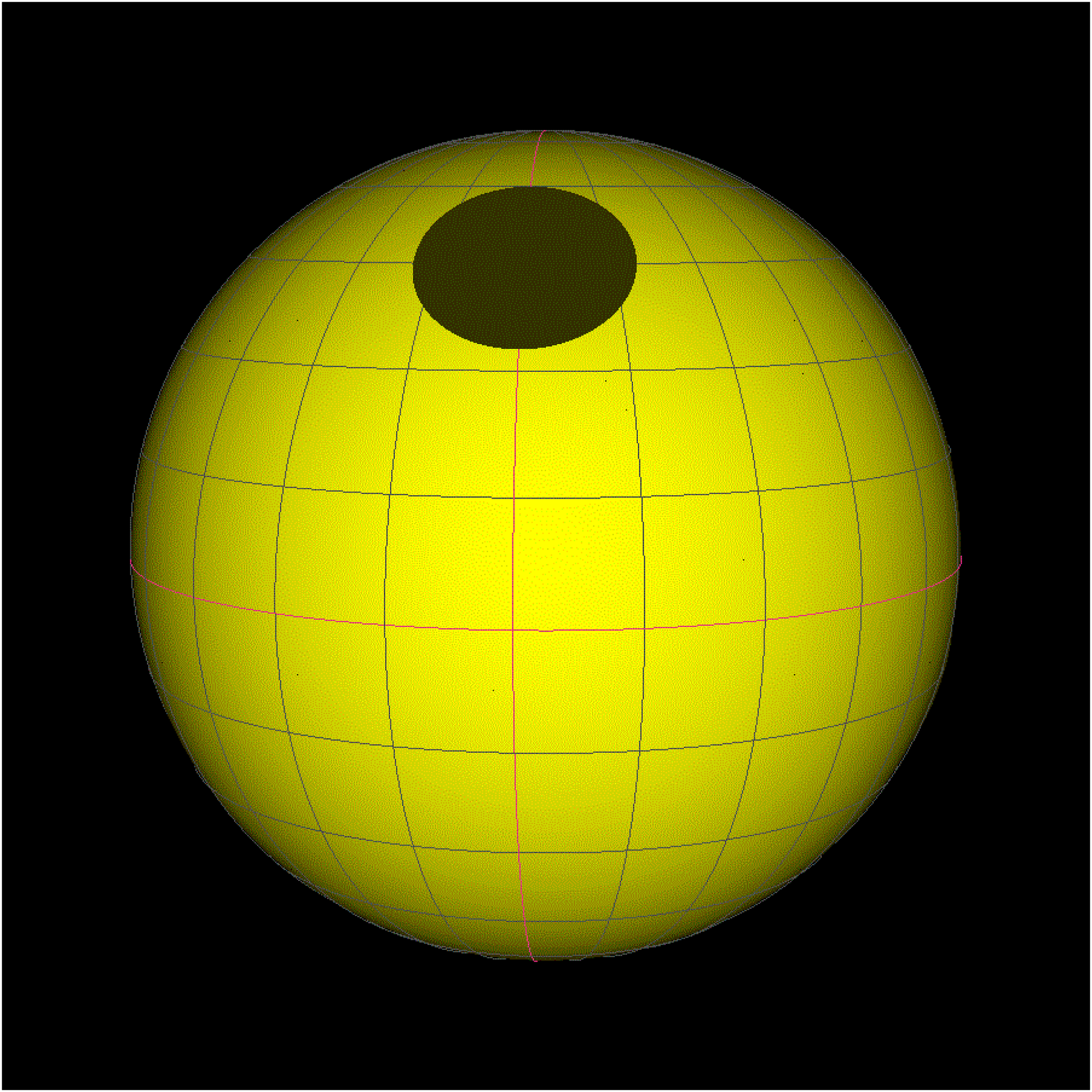}
\includegraphics[scale = 0.045, angle = 0]{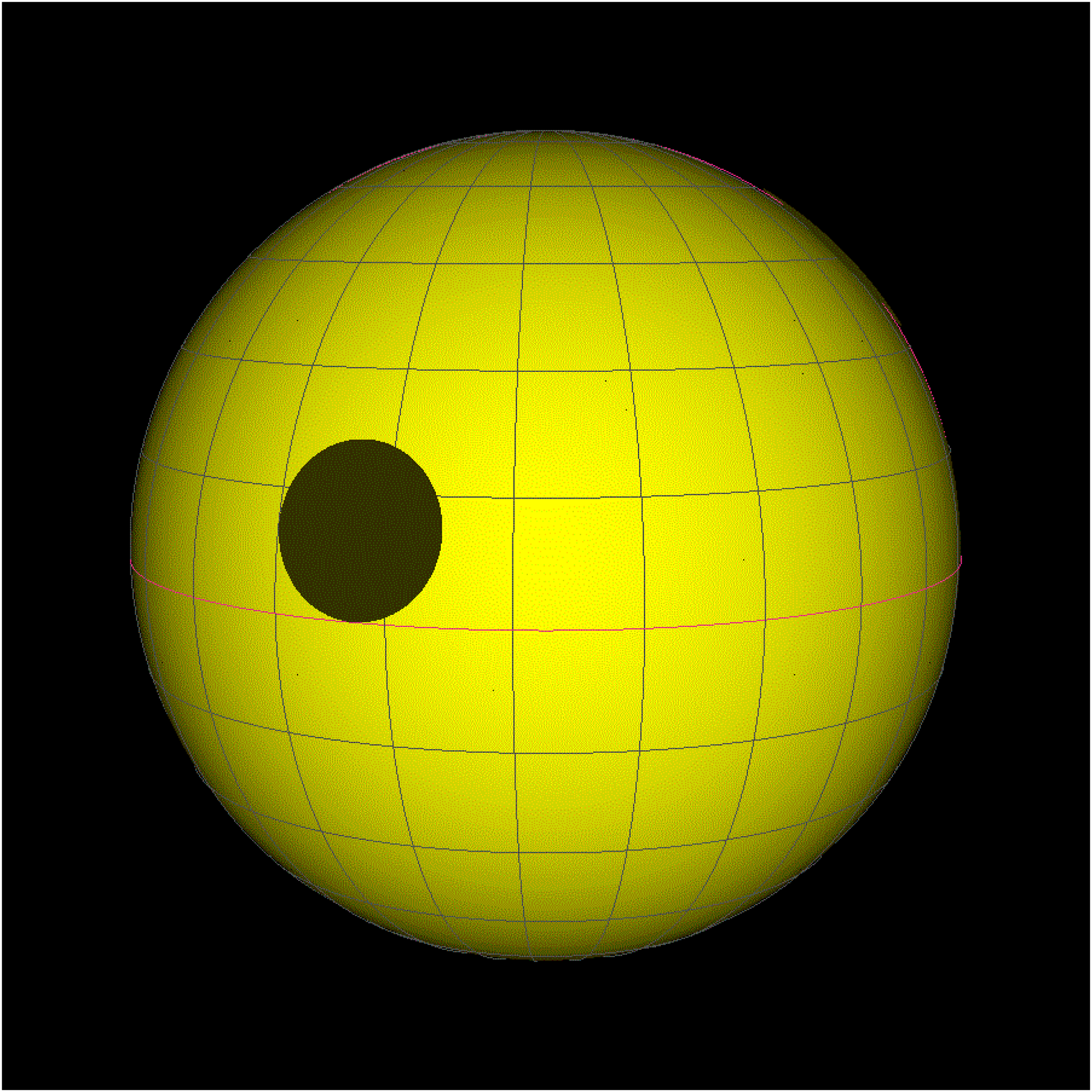}
\includegraphics[scale = 0.045, angle = 0]{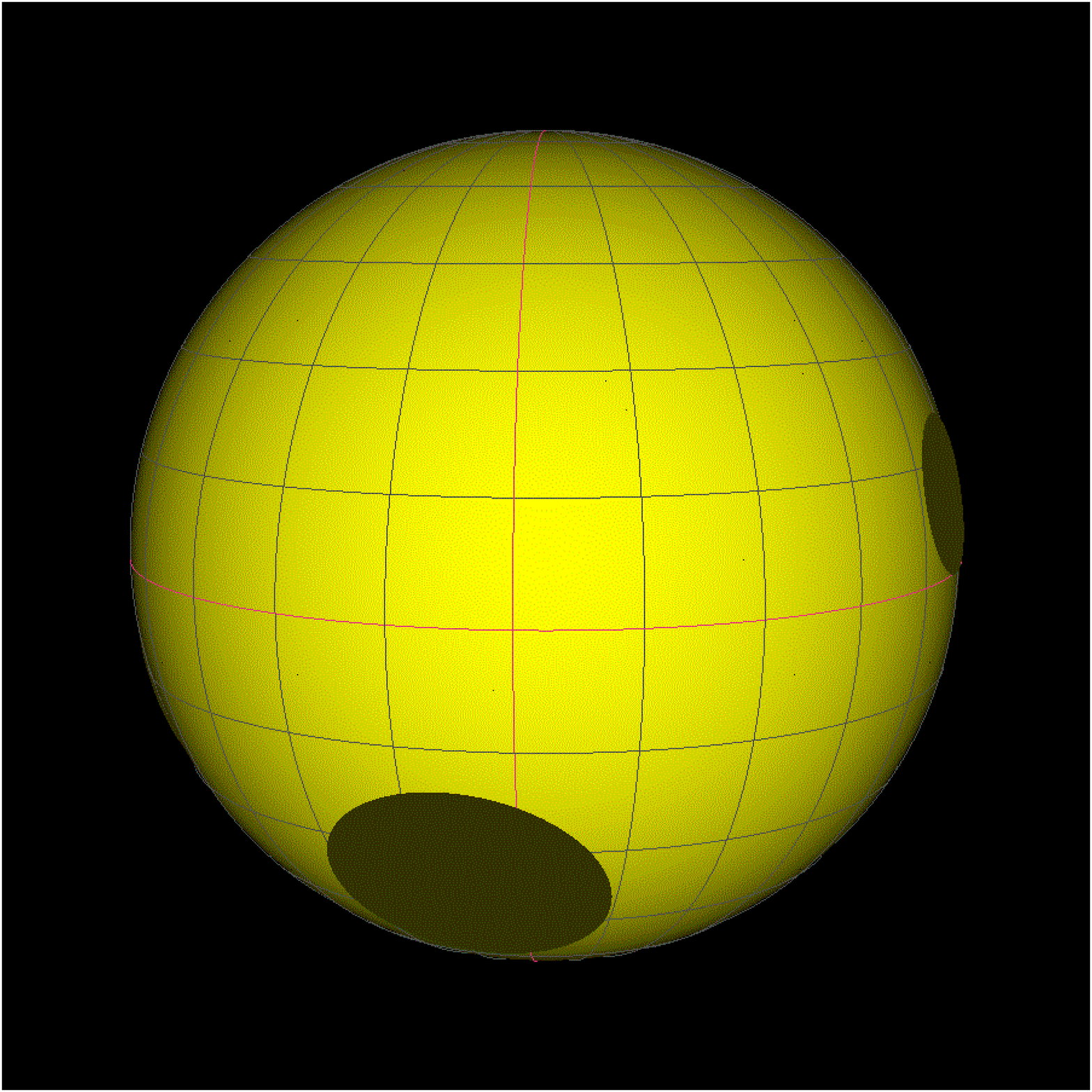}
\includegraphics[scale = 0.045, angle = 0]{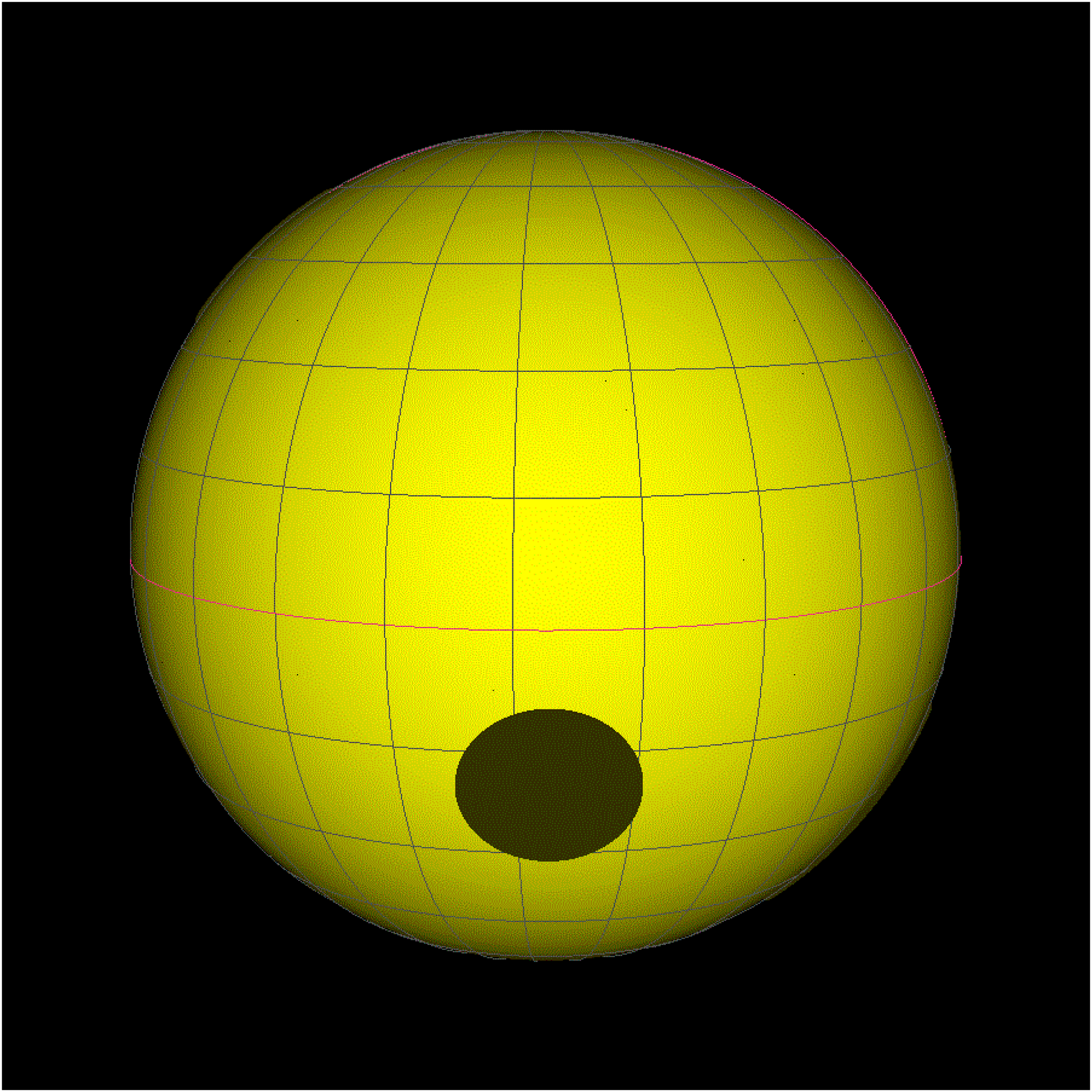}
\includegraphics[scale = 0.46, angle = 270]{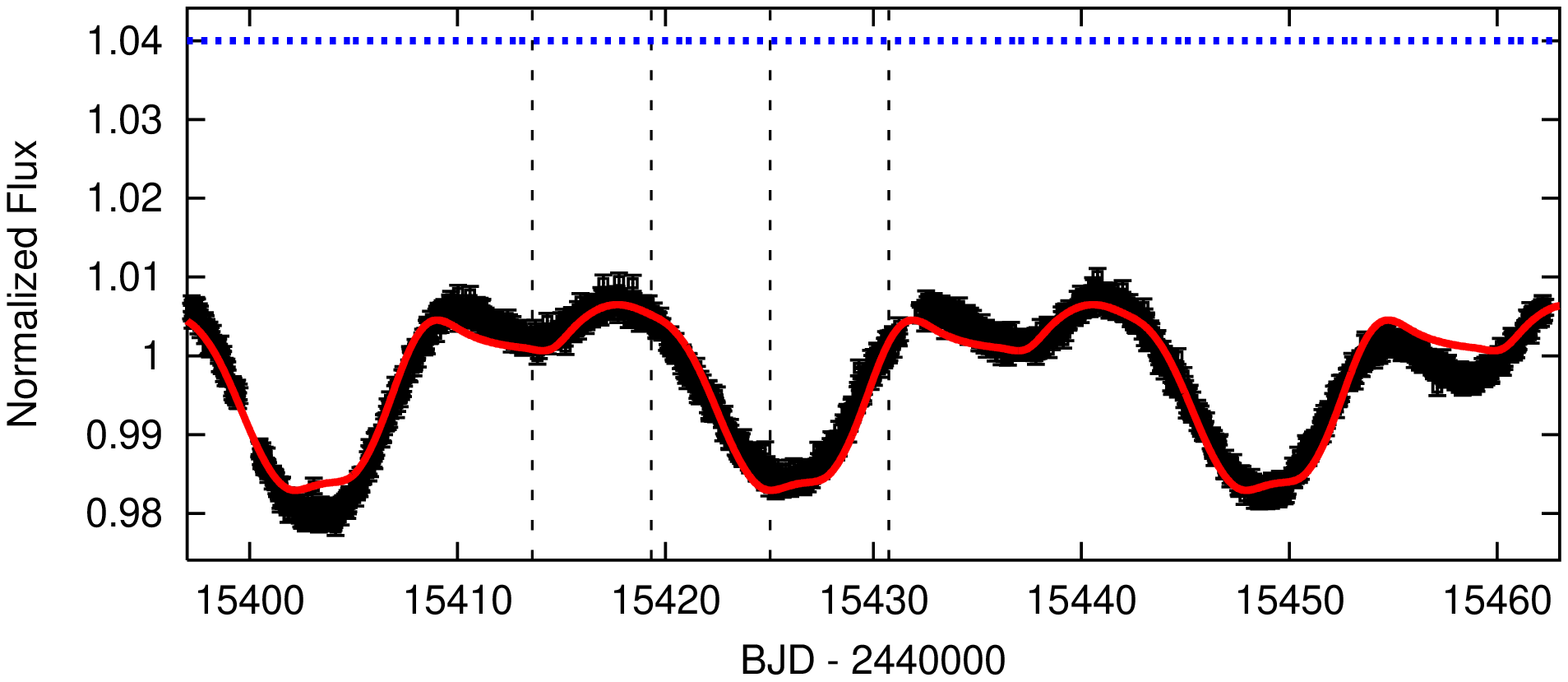}
\caption[]
        {Top panels: Possible spot model of the rotational modulation of KIC 1255
	as seen from the line of sight at stellar rotation phases $\theta$=0.15, 0.40, 0.65 and 0.90 (from left).
	The bottom panel displays our {\it StarSpotz} fit (the red solid curve) to
	the {\it Kepler} long cadence photometry (black points with error bars) for a subset of the data.
	The blue horizontal dotted line displays the unspotted photosphere of the star.
	The vertical dotted black lines indicate phases 0.15, 0.40, 0.65 and 0.90 (from left).
	If the cometary tail of KIC 1255b transited the spots near the top of the visible portion of the star,
	this would cause sufficiently shallow transits to explain the KIC 1255 TDRM effect, despite
	the fact these spots only appear to cause 1\% drops below the maximum flux level observed.
}
\label{FigStarSpotz}
\end{figure}

To illustrate that what appears to be a 1\% flux drop could actually be caused by 
a much larger spot, we
fit a section of the KIC 1255 {\it Kepler} long cadence light curve with a \citet{Budding77} 
model using the techniques
of {\it StarSpotz} \citep{CrollEpsEri,CrollMCMC,Walker07}.
We fit a subset of the long cadence {\it Kepler} photometry from BJD=2455397.2 to BJD=2455462.3.
We use values for the flux ratio
of the spotted to unspotted photosphere, $I_s/I_*$=0.2,
and the linear limb-darkening coefficient, $u$=0.8, similar to what were employed
previously for another K-dwarf: $\epsilon$ Eridani \citep{CrollEpsEri}.
We set the normalized flux of the unspotted photosphere as $U$ = 1.04, an inclination angle
to the line-of-sight of $i$=80$^o$,
and place four starspots on the star.
We display our possible spot model in Figure \ref{FigStarSpotz}.
The spot near the top of the star (the top visible portion of the star) causes a
drop in flux of approximately 4\% of the flux of the star.
If this spot were occulted by the cometary tail trailing KIC 1255b,
the effect would be large enough to explain the KIC 1255 TDRM signal.
This is not in any way intended as a unique model to explain the observed 
rotational modulation of KIC 1255, but simply illustrative that our occulted spot explanation is plausible
despite the shallowest transit depths being observed at a stellar rotational phase of $\theta$=0.2.
There are likely a variety of other spot combinations that could serve to satisfy these conditions
and could explain the KIC 1255 TDRM signal.

\subsection{Model with Periodically Enhanced Mass Loss}

\begin{figure}
\centering
\includegraphics[scale = 0.46, angle = 0]{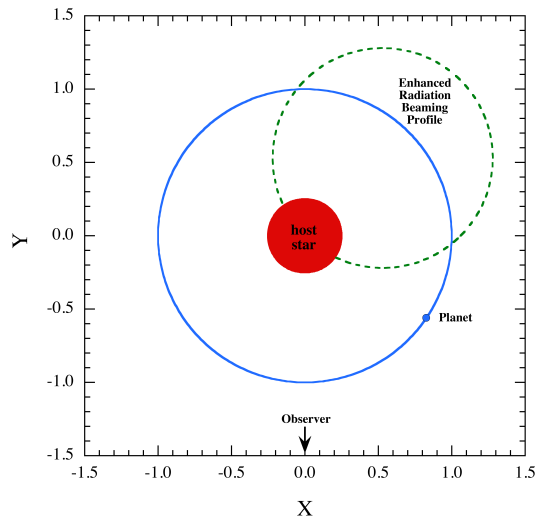}
\caption[]
        {
	Schematic in the orbital plane of the KIC 1255 system with a
	hypothetical region of enhanced radiation through which the planet passes
	periodically.  The green dotted curve represents the beaming function and
	is assumed to have a simple Lambertian profile.  This is shown
	graphically as the distance from the origin (the middle of the host star) to the green curve as a
	function of angle from the stellar longitude where the enhancement peaks.
	The model discussed in the text assumes that the excess mass loss rate
	is proportional to the intensity of radiation in this region.  Note that
	in this simple model there will be some enhancement in the mass
	loss rate over essentially half the orbit.  The X and Y coordinates are
	normalized to the orbital radius of the planet.
}
\label{FigKawaharaSchematic}
\end{figure}

%EMULATEAPJCHANGE, 0.55 to 0.50
\begin{figure*}
\centering
\includegraphics[scale = 0.55, angle = 0]{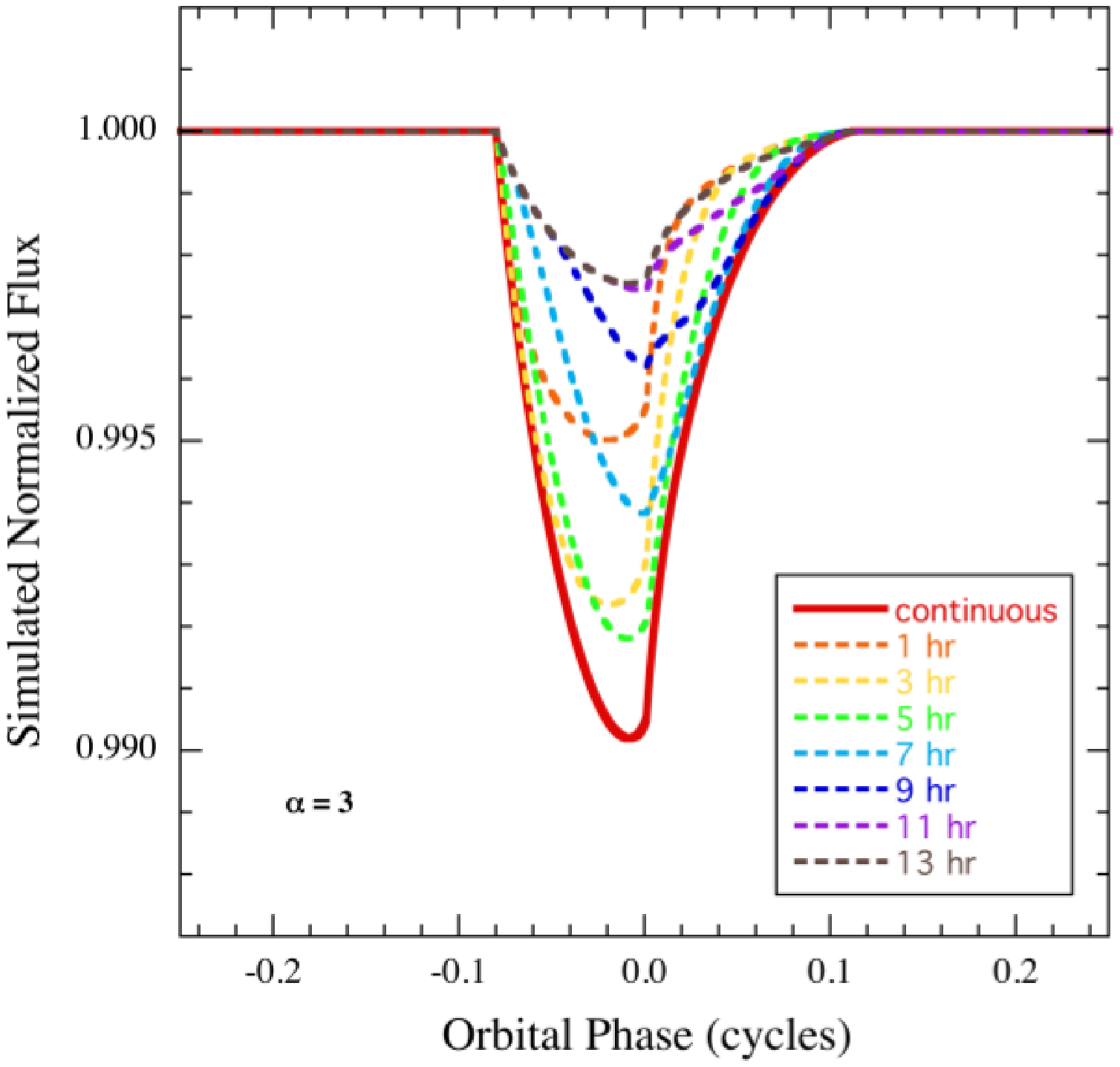}
\includegraphics[scale = 0.55, angle = 0]{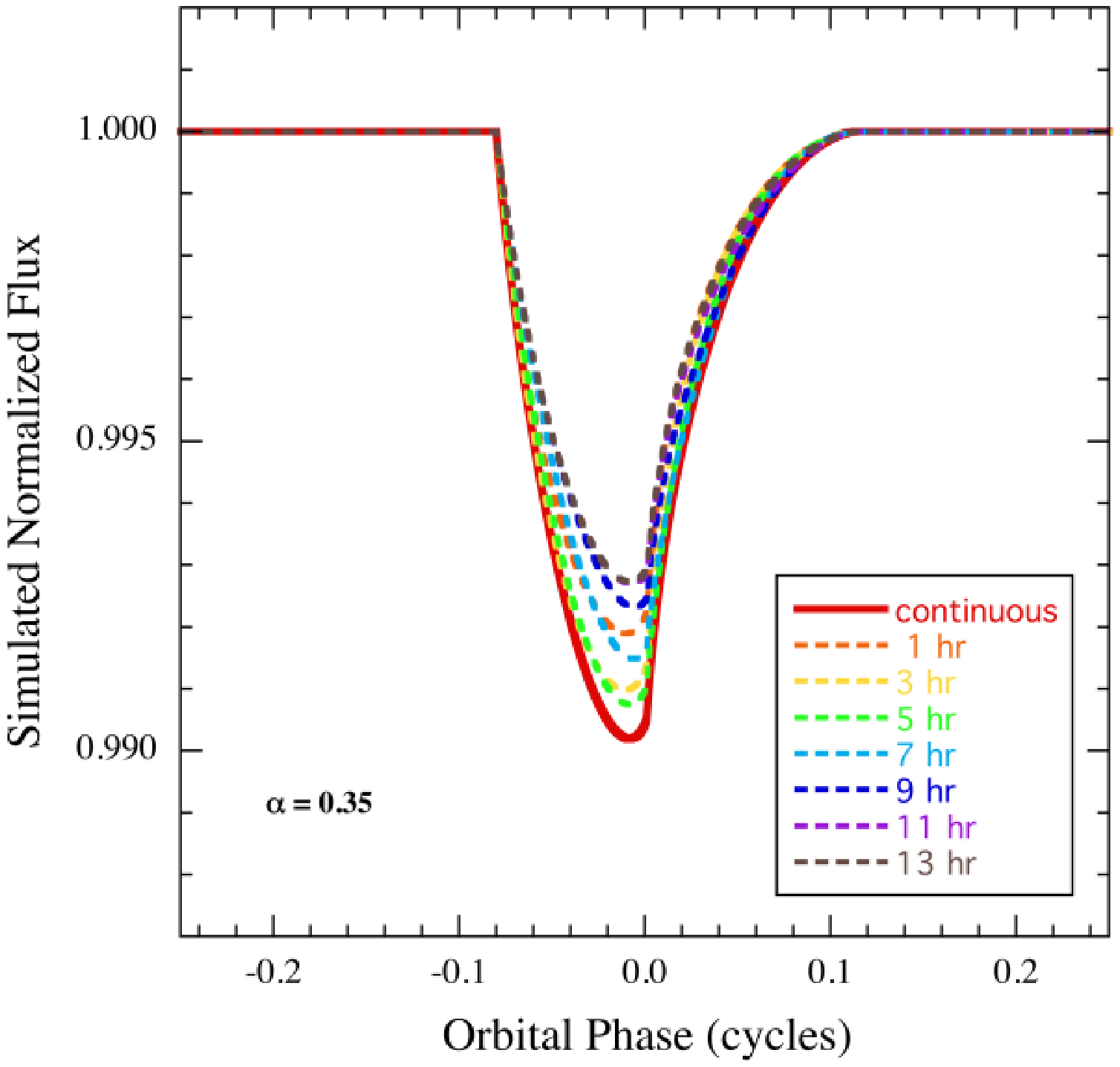}
\caption[]
        {
        Dust-tail transit profiles for KIC 1255b calculated on the basis of the toy model discussed in the text.
        In this model, excess mass loss rates are stimulated each orbit when the planet passes through a beam of enhanced
        radiation which corotates with the host star.  The color coding indicates the amount of time between passage through
        the center of the enhanced radiation field and the time to the following transit.  After the dust-density profiles are computed (see
        text for details), we convolve the transmission profile with a limb-darkened stellar disk assuming an equatorial transit.  
        The relative opacity of the dust is scaled until the resultant transit depth is about 1\%.  Left panel: the case where the 
        mass loss rate at the peak of the enhanced radiation is a factor of 1+$\alpha$=4 times the steady state from
	the rest of the star.  Right panel: the case 
        where the enhanced mass loss rate is a factor of 1+$\alpha$=1.35 times the rest of the star. 
        }
\label{FigTail}
\end{figure*}

 Inspired by the \citet{Kawahara13} explanation for the TDRM signal -- of an active longitude of enhanced
ultraviolet or X-ray radiation or some other sort of star-planet interaction --
in this section we attempt to evaluate the plausibility of a model wherein KIC 1255b passes through a region
that drives enhanced mass loss from the planet. That is, we assume that once 
per synodic orbital period, the planet passes through an enhanced radiation field, leading to an increase
in the mass loss rate in dust from the planet. A schematic of the scenario is shown in Figure \ref{FigKawaharaSchematic}.  We make no
attempt to describe the origin of this amplified radiation field or the details of how this leads to enhanced
mass loss. 

% In order to show what such an effect might have on the dust tail during transits, and how it would
% affect the observations, we consider the extreme (i.e., illustrative only) case where there is no mass loss
% from the planet except for times that it is immersed in the enhanced radiation field. This cannot, in fact, be
% the case since we see dramatic changes in transit depths from transit to transit, and the effect we expect from
% the model being discussed would occur on the rotation period of the host star (which is $\sim$35 times longer
% than the orbital period).  Nonetheless, the extreme case is quite instructive.

There are necessarily many things that we do not know about this hypothetical process
of enhanced radiation
from a specific longitude on the star
leading to an increase in the mass loss rate from the planet, but we will choose an
illustrative set of parameters that can hopefully show how such an effect might work.  For one, the details of
how radiation at various wavelengths interacts with the planet, heats the surface, and drives off heavy metal
vapors are highly uncertain; they are discussed at length by 
\citeauthor{PerezBeckerChiang13} (\citeyear{PerezBeckerChiang13}; see also \citealt{Rappaport12}).
The dynamics of the dust, once it condenses from the heavy metal vapor during the
escape from the planet, are discussed in some detail by \citet{Rappaport14}
in the context of a second candidate
``evaporating'' planet (KOI-2700b).

In this simplistic toy model, which describes the effects of periodically enhanced mass loss, we assume that
(1) $\dot M_{\rm dust}$ is directly enhanced by the instantaneous excess radiation field at the planet; (2)
the radiation enhancement is described by a simple Lambertian beaming profile in orbital phase and
centered at a particular stellar longitude
leading to 
an enhanced mass loss rate that at its peak is a value $\alpha$ above the steady state value (the steady state value is set to be unity);
(3) there is a period of constant coasting velocity for the dust away
from the planet at several times the escape speed until the dust has reached several planetary radii; (4) the
dust is subsequently subjected to only radiation pressure and to the gravity of the host star (with a ratio
$\beta$ for these two effects); and (5) the effective scattering cross section of the dust decays exponentially
in time with a time-constant, $\tau_{\rm dust}$, due to sublimation.  The longitude where the enhanced radiation
is centered (see Figure \ref{FigKawaharaSchematic}) is assumed to corotate with the host star every $P_{\rm rot} =22.9$ days.

For each assumed longitude where the enhanced radiation is centered, we compute how far the dust has moved
down the tail by the time of the next planetary transit.
The motion of the dust particles in the tail, relative to the planet, was determined by integrating Equation (3) of
\citet{Rappaport14}.  The density of a parcel of dust is taken to be proportional
to the radiation field at the planet at the time the dust was emitted, and to have decayed exponentially in time.
The two key parameters in this ad hoc model are the ratio of radiation
pressure to gravity, $\beta$, which we arbitrarily take to be 0.07 (see Appendix B of \citealt{Rappaport14}),
and the exponential decay time for the effective cross section of a dust grain, $\tau_{dust}$, which we arbitrarily take to be
2.5 hours.  The dust extinction at that location in the tail is taken to be proportional to the dust density, and the
attenuation profile is thereby calculated.  Finally, we move the longitude of the centroid of the enhanced
radiation beam (as a proxy for the rotating host star), and repeat the calculation.

The results are presented in Figure \ref{FigTail}.  We show a series of calculated transit profiles for
different time intervals between the planet's passage through the enhanced radiation zone and the time of the
next transit.  To produce these curves, we have taken each of the simulated dust-density profiles and convolved 
it with a limb-darkened 
star\footnote{Again using a linear limb-darkened coefficient of $u$=0.8.}. 
The two cases shown are for a peak mass loss rate 
that is a factor of 1+$\alpha$=4 times the steady state, unenhanced mass loss rate 
from the rest of the star (left panel), 
and another that is 1+$\alpha$=1.35 times the steady state (right panel).
The heavy red curve in each plot is the expected transmission profile, using the same model, except that the 
enhanced radiation zone is made so large as to produce a nearly continuous outflow of dust.
The left panel of Figure \ref{FigTail}, especially, indicates that not
only does the proposed mechanism cause transit-depth variations, it can also lead
to transit-timing and transit-profile variations.
We note that our $\alpha$=0.35 model produces transit depth
variations on the order of 27\%, similar to the KIC 1255 TDRM signal.

We conclude from these calculations that such a hypothetical model of enhanced driven mass loss, at a 
preferential stellar longitude that rotates with the host star, can quite naturally produce systematic changes 
in the transit depth, of the correct magnitude and in phase with the 22.9-day rotation period. For our 1.35-times enhanced
mass loss model, the shapes
of the transit profiles do not change dramatically with the 22.9-day orbital phase.  This is consistent 
with the fact that no obvious transit profile changes were visible in the KIC 1255 {\it Kepler} long cadence 
photometry (Figure \ref{FigTransitProfile}).  We note that in spite of the lack of predicted dramatic changes in 
transit profile, there might be imperceptible changes that can still lead to measurable transit-timing variations --- 
and we can check this with our model.
After convolving our profiles with the 
{\it Kepler} long cadence integration time, we subjected our profiles
to a transit timing analysis.
Our model with a mass-loss enhancement of 1.35-times the steady-state
($\alpha$=0.35; right panel in Figure \ref{FigTail}),
produces transit depth variations on the order of $\sim$27\%, 
and transit-timing variations on the order of $\sim$\TransitTimingVariationsKawaharaSeconds \ s.
This value is marginally above our 
stringent 3$\sigma$ limit of \TransitTimingLimitTwiceSeconds \ s on the 
peak-to-peak timing variations on the transit of KIC 1255b phased to the stellar rotation period for all the data
(Figure \ref{FigTiming}).
If we only analyze the first half
of the {\it Kepler} long cadence data (excluding the quiescent periods; BJD - 2440000 = 15078 - 15600),
where the TDRM signal is stronger (Figure \ref{FigKawaharaSplit}), such transit-timing variations are not ruled out.
Although the \citet{Kawahara13} proposed mechanism would naturally lead to transit-timing variations,
our analysis indicates that these transit-timing
variations could be small enough that they could have escaped detection in the 
{\it Kepler} long cadence data.

We acknowledge that there are a number of parameters associated with our toy model that one could quibble with; nonetheless,
the above simulation and `analysis' suggests that it is possible that an active longitude could
drive enhanced mass loss and transits 
that are 25\% deeper at one stellar rotation phase than another, with transit-timing variations
that are small enough that they would not have been ruled out by the {\it Kepler} long cadence photometry of KIC 1255.
We therefore cannot rule out the validity of the model proposed by \citep{Kawahara13}.

\subsubsection{Validity of the ultraviolet flux as a driver of an enhanced mass loss rate}

The mechanism most discussed for removing gaseous heavy metal material from
KIC 1255b has been a Parker-type wind. This wind is generated by heating of the base
of the atmosphere by the broadband radiation from the parent star which
drives a hydrodynamic outflow \citep{Rappaport12,PerezBeckerChiang13}.
It is also possible that the mass loss rate could be supplemented,
or even dominated, by photoevaporation of the planetary atmosphere by the X-ray
and/or extreme ultraviolet (EUV)
flux from the host star as alluded to in the model proposed by \citet{Kawahara13}.
We can utilize the prescription of \citet{SanzForcada11} for
estimating the nominal EUV luminosity of KIC 1255:
$$\log(L_{\rm EUV}) \simeq 29.1 -1.2\log(\tau)$$
where $\tau$ is the age of the star in Gyr, and we adopt a gyrochronological age of
1 Gyr \citep{Barnes07}.  If we further assume that the EUV flux dominates over the X-ray flux \citep{SanzForcada11},
we can utilize the expression for photoevaporative mass loss rates following
\citet{Watson81}, \citet{Lammer03}, and \citet{SanzForcada11}, to write
\begin{eqnarray*}
\label{eqn:nevap}
\dot M_{\rm evap} \simeq 6 \times 10^{11} \left(\frac{F_{\rm EUV}} {2 \times 10^5 \,{ergs \ sec^{-1} \ cm^{-2}}}\right) \left( \frac{5 {g \ cm^{-3}}}{\rho_p} \right) {\rm g\,s}^{-1}
\end{eqnarray*}
where $\rho_p$ is the mean density of the planet.  This is sufficient from an energetics point
of view to eject the inferred mass loss rates from KIC 1255b \citep{Rappaport12}.
We might therefore conclude
that a persistent longitude on the host star with excess EUV emission might plausibly account
for the TDRM signal.
However, this mechanism is likely to be
much more efficient at removing lighter, higher-velocity molecules -- such as H, He, and H$_2$O --
than the heavy metal molecules that would be needed to form
dust from the atmosphere of a rocky planet.
Thus, it may be that photoevaporation of rocky planets may not be nearly so efficient as it
could be for gas giants, therefore calling into question the validity of this model in regard to the
supposedly sub-Mercury-mass
KIC 1255b \citep{CrollKIC}.

\subsubsection{GALEX observations of KIC 1255}

Another way to constrain the \citet{Kawahara13} model that there is an active longitude on KIC 1255 that is displaying
enhanced ultraviolet or X-ray emission is to actually observe KIC 1255 at these wavelengths to see
if the flux is modulated at the rotation period of the star. 
The {\it Galaxy Evolution Explorer} (GALEX; \citealt{Martin03}) performed ultraviolet observations of the {\it Kepler}-field,
but these observations were not able to detect KIC 1255; 
there is no near-ultraviolet source at the position of KIC 1255 to a magnitude limit of $m_{AB}$ $<$ 22 mag.
As KIC 1255 is relatively faint, this does not rule out a moderately young, active star (Jamie Lloyd
\& Everett Schlawin, private communication).

% Many of the details of such a model, not to mention the unknown origin
% of the fixed region of enhanced radiation, remain to be worked out.

\section{Summary}

In this work we attempted to confirm the claimed signal of \citet{Kawahara13}
wherein the depth of the transits of KIC 1255b were correlated with the phase of the stellar rotation.
%We refer to this Transit Depth Rotational Modulation signal as the TDRM signal.
%The main conclusions of our work are as follows:
We confirmed that there is indeed a robust, statistically significant correlation between
the depths of KIC 1255b's transits and the stellar rotation period.
This signal is not due to 
the leakage of the rotating spot signal into our measurement of the transit depths,
or due to unocculted starspots.
The transits of KIC 1255b are approximately 25\% deeper at one
stellar rotation phase than another.
We show that the effect is stronger in the first-half of the {\it Kepler} light curve than 
in the second-half.
We also perform a transit timing analysis of
the transits of KIC 1255b and are 
able to place a 3$\sigma$ upper-limit on the peak to peak amplitude of these variations 
phased to the stellar rotation period of
$\sim$\TransitTimingLimitTwiceSeconds \ s.

To help us understand the 
transit depth rotational modulation signal, 
as KIC 1255 is a very spotted star with rapidly evolving spots,
we also searched for 
such signals in the {\it Kepler} light curves of several other spotted stars with transiting
planets.
A transit depth rotational modulation signal is observed in 
two other spotted stars with transiting planets that feature 
starspot occultations by the transiting planet.
Due to the unique geometry of the candidate disintegrating planet KIC 1255b,
if its dust tail occults a cool starspot, we show that the anomalous brightening during transit due to the occultation
will persist for a longer fraction of the transit than for other transiting planets.
A likely explanation for this signal is that the dust tail trailing KIC 1255b
occults starspots, leading to shallower transits at certain stellar rotation phases.
Such a model could lead to large transit-timing variations (up to $\sim$\TransitTimingVariationsOccultedSpotsSeconds \ s); however,
if the planet is occulting a number of small spots, the associated transit-timing variations could be much smaller,
and therefore we feel this model is arguably consistent with our upper limit on timing variations of the transits of KIC 1255b.

 We also investigate the suggestion that the transit depth rotational modulation signal could be due to
an active longitude on the star that causes an increased mass loss rate from the planet; we employ a toy model that suggests
that such a model could naturally lead to transit depth variations similar to what we observe. This model
could similarly lead to transit-timing variations ($\sim$\TransitTimingVariationsKawaharaSeconds \ s variations), although such variations
could have been small enough that they escaped detection in the {\it Kepler} long cadence photometry. 

 For these reasons we believe that both occulted spots and an active region driving increased mass loss remain viable options to
explain the Transit Depth Rotational Modulation signal. 
However, as there is not currently any empirical 
evidence that there is an active region on KIC 1255 that is exhibiting enhanced
ultraviolet or X-ray flux, we believe the occulted spot scenario provides the simplest 
explanation for the observations.
Therefore, our preferred explanation 
for the statistically significant correlation between the 
KIC 1255b transit depths and the phase of the stellar rotation period is that the cometary tail trailing KIC 1255b
occults cool starspots, leading to slightly shallower transit depths during these occultations.

\acknowledgements
B.C.'s work was performed under a contract with the California Institute of
Technology funded by NASA through the Sagan Fellowship Program.
We thank Roberto Sanchis-Ojeda \& Kevin Schlaufman for helpful discussions that contributed to this work.
We thank Jamie Lloyd \& Everett Schlawin for sharing their analysis of the GALEX observations of KIC 1255.


\begin{thebibliography}{}

\bibitem[Bakos et al. (2010)]{Bakos10} Bakos, G.A. et al. 2010, \apj, 710, 1724

\bibitem[Barnes (2007)]{Barnes07} Barnes, S.A. 2007, \apj, 669, 167

\bibitem[Barros et al. (2013)]{Barros13} Barros, S.C.C. et al. 2013, \mnras, 430, 3032

\bibitem[Borucki et al. (2009)]{Borucki09} Borucki, W.J. et al. 2009, Science, 325, 709

\bibitem[Brogi et al. (2012)]{Brogi12} Brogi, M. et al. 2012, \aap, 545, L5

\bibitem[Budaj (2013)]{Budaj13} Budaj, J. 2013, \aap, 557, A72

\bibitem[Budding (1977)]{Budding77} Budding, E. 1977, Ap\&SS, 48, 207

\bibitem[Carter et al.(2011)]{Carter11} Carter, J.A. et al. 2011, \apj, 730, 82

\bibitem[Croll et al. (2006)]{CrollEpsEri} Croll, B. et al. 2006, \apj, 648, 607

\bibitem[Croll (2006)]{CrollMCMC} Croll, B. 2006, \pasp, 118, 1351

\bibitem[Croll et al.(2014)]{CrollKIC} Croll, B. et al. 2014, \apj, 786, 100

\bibitem[Czela et al.(2009)]{Czela09} Czela, S. et al. 2009, \aap, 505, 1277

\bibitem[Deming et al.(2011)]{Deming11} Deming, D. et al. 2011, \apj, 740, 33

\bibitem[Desert et al.(2011)]{Desert11} Desert, J-M. et al. 2011, \apjs, 197, 14

\bibitem[Kawahara et al. (2013)]{Kawahara13} Kawahara, H. et al. 2013, \apjl, 776, L6

\bibitem[Koch et al. (2011)]{Koch10} Koch, D.~G., Borucki, W.~J et al. 2010, \apjl, 713, L79

\bibitem[Lammer et al. (2003)]{Lammer03} Lammer, H. et al. 2003, \apjl, 598, L121

\bibitem[Lomb (1976)]{Lomb76} Lomb, N.~R. 1976, Ap\&SS, 39, 447

\bibitem[Martin et al. (2003)]{Martin03} Martin, C. et al. 2003, in Society of Photo-Optical Instrumentation Engineers (SPIE) Conference Series, Vol. 4854, Society of Photo-Optical Instrumentation Engineers (SPIE) Conference Series

\bibitem[Perez-Becker \& Chiang (2013)]{PerezBeckerChiang13} Perez-Becker, D. \& Chiang, E. 2013, \mnras, 433, 2294

\bibitem[Rappaport et al. (2012)]{Rappaport12} Rappaport, S. et al. 2012, \apj, 752, 1

\bibitem[Rappaport et al. (2014)]{Rappaport14} Rappaport, S. et al. 2014, \apj, 2014, 784, 40

\bibitem[Sanchis-Ojeda \& Winn (2011)]{SanchisOjeda11} Sanchis-Ojeda, R. \& Winn, J.N. 2011, \apj, 743, 61

\bibitem[Sanchis-Ojeda et al. (2013)]{SanchisOjeda13} Sanchis-Ojeda, R. et al. 2013, \apj, 774, 54

\bibitem[Sanz-Forcada et al. (2011)]{SanzForcada11} Sanz-Forcada, J. et al. 2011, \aap, 532, A6

\bibitem[Scargle (1982)]{Scargle82} Scargle, J.D. 1982, \apj, 263, 835

\bibitem[Smith et al. (2012)]{Smith12} Smith, J.C. et al. 2012, \pasp, 124, 1000

\bibitem[Stumpe et al. (2012)]{Stumpe12} Stumpe, M.C. et al. 2012, \pasp, 124, 985

\bibitem[van Werkhoven et al. (2014)]{vanWerkhoven14} van Werkhoven, T.I.M. et al. 2014, \aap, 561, A3

\bibitem[Walker et al. (2007)]{Walker07} Walker, G.A.H. et al. 2007, \apj, 659, 1611

\bibitem[Watson et al. (1981)]{Watson81} Watson, A.J. et al. 1981, Icar, 48, 150

\end{thebibliography}
\end{document}